\documentclass[preprints,article,submil,pdftex,oneauthor]{Definitions/mdpi} 

\firstpage{1} 
\pubvolume{1}
\issuenum{1}
\articlenumber{0}
\pubyear{2023}
\copyrightyear{2023}
\externaleditor{Academic Editor(s): Ivanka Stamova, Gani Stamov}
\datereceived{4 September 2023} 
\daterevised{27 September 2023}
\dateaccepted{28 September 2023} 
\datepublished{ } 
\hreflink{https://doi.org/} 

\usepackage{subcaption}
\usepackage{xfrac}
\usepackage{amsmath}
\DeclareMathOperator{\sign}{sign}
\newcommand{\Mathematica}{\textit{Mathematica\textsuperscript{\resizebox{!}{0.8ex}{\textregistered}}}}
\def\8{\infty}
\def\oh{\sfrac{1}{2}}
\def\ot{\sfrac{1}{3}}
\def\oq{\sfrac{1}{4}}
\def\tt{\sfrac{2}{3}}

\newcommand*{\I}{\imath}%

\def\eps{\epsilon}

\def\dal{\partial_{\alpha}}
\def\dbe{\partial_{\beta}}

\def\const{\textit{ const }}
\def\undertext#1{\vtop{\hbox{#1}\kern 1pt \hrule}}
\def\Ra{\Rightarrow}

\def\abs#1{\left| #1\right|}

\def\VEV#1{\left\langle #1\right\rangle}

\def\tr{\hbox{tr}\,}

\def\pp#1{\frac{\partial}{\partial#1}}
\def\pbyp#1#2{\frac{\partial#1}{\partial#2}}
\def\ff#1{\frac{\delta}{\delta#1}}
\def\fbyf#1#2{\frac{\delta#1}{\delta#2}}

\def\inv#1{\frac{1}{#1}}
\def\bea{\begin{eqnarray} & &}
\def\eea{\end{eqnarray}}

\let\oldexp\exp
\renewcommand{\exp}[1]{\oldexp\left(#1\right)}

\def\IINT#1{\int_{#1}}

\def\CL{Clebsch}

\def\NS{Navier-Stokes}
\def\KH{Kelvin-Helmholtz}
\def\BS{Biot-Savart}

\def\SYM {symplectomorphisms}

\def\val{v_{\alpha}}
\def\vbe{v_{\beta}}

\def\ral{r_{\alpha}}
\def\rbe{r_{\beta}}

\newcommand{\Mod}[1]{\ (\mathrm{mod}\ #1)}

\def\XXint#1#2#3{{\setbox0=\hbox{$#1{#2#3}{\int}$}
     \vcenter{\hbox{$#2#3$}}\kern-.5\wd0}}

\newcommand{\bS}{\mathbb{S}}

\newcommand{\bC}{\mathbb{C}}

\renewcommand{\Re}{\textbf{Re }}
\renewcommand{\Im}{\textbf{Im }}
\def\Binom#1#2{\dbinom{#1}{#2}}

\newcommand{\tpmod}[1]{{\@displayfalse\Mod{#1}}}
\usepackage{mathtools}

\DeclarePairedDelimiter\floor{\lfloor}{\rfloor}
\newcommand{\pct}[2]{
\begin{figure}
    \includegraphics[width=\textwidth]{#1}
    \caption{#2}
    \label{fig::#1}
\end{figure}
}

\Title{TO THE THEORY OF DECAYING TURBULENCE}

\TitleCitation{Title}

\Author{Alexander Migdal $^{1,\dagger,\ddagger}$\orcidA{}}

\AuthorNames{Alexander Migdal}

\AuthorCitation{Migdal, A.~A.}

\address{
$^{1}$ \quad Department of Physics, New York University Abu Dhabi,
            Saadiyat Island, 
            Abu Dhabi,
            PO Box 129188, 
           Abu Dhabi,
            United Arab Emirates; am10485@nyu.edu}

\corres{Correspondence: sasha.migdal@gmail.com}

\abstract{We have found an infinite dimensional manifold of exact solutions of the Navier-Stokes loop equation for the Wilson loop in decaying Turbulence in arbitrary dimension $d >2$. This solution family is equivalent to a fractal curve in complex space $\mathbb C^d$ with random steps parametrized by $N$ Ising variables $\sigma_i=\pm 1$, in addition to a rational number $\frac{p}{q}$ and an integer winding number $r$, related by $\sum \sigma_i = q r$. This equivalence provides a dual theory describing a strong turbulent phase of the Navier-Stokes flow in $\mathbb R_d$ space as a random geometry in a different space, like ADS/CFT correspondence in gauge theory. From a mathematical point of view, this theory implements a stochastic solution of the unforced Navier-Stokes equations. For a theoretical physicist, this is a quantum statistical system with integer-valued parameters, satisfying some number theory constraints. Its long-range interaction leads to critical phenomena when its size $N \rightarrow \infty$ or its chemical potential $\mu \rightarrow 0$. The system with fixed $N$ has different asymptotics at odd and even $N\rightarrow \infty$, but the limit $\mu \rightarrow 0$ is well defined. The energy dissipation rate is analytically calculated as a function of $\mu$ using methods of number theory. It grows as $\nu/\mu^2$ in the continuum limit $\mu \rightarrow 0$, leading to anomalous dissipation at $\mu \propto \sqrt{\nu} \to 0$. The same method is used to compute all the local vorticity distribution, which has no continuum limit but is renormalizable in the sense that infinities can be absorbed into the redefinition of the parameters. The small perturbation of the fixed manifold satisfies the linear equation we solved in a general form. This perturbation decays as $t^{-\lambda}$, with a continuous spectrum of indexes $\lambda$ in the local limit $\mu \to 0$.
The spectrum is determined by a resolvent, which is represented as an infinite product of $3\otimes3$ matrices depending of the element of the Euler ensemble.}


\begin{document}

\section{Introduction 
}

We derived a functional equation for the so-called loop average or Wilson loop in turbulence a while ago. All the references to our previous works can be found in a recent review paper \cite{M23PR}.

The path to an exact solution by a dimensional reduction in this equation was proposed in the 1993 paper but has just been explored.
At the time, we could not compare a theory with anything but crude measurements in physical and numerical experiments at modest Reynolds numbers.
All these experiments agreed with the K41 scaling, so the exotic equation based on unjustified methods of quantum field theory was premature.
The specific prediction of the Loop equation, namely the Area law, could not be verified in DNS at the time with existing computer power.

\subsection{Fluctuating geometry in gauge theory and turbulent flow}
The situation has changed over the last decades. No alternative microscopic theory based on the \NS{} equation emerged, but our understanding of the strong turbulence phenomena grew significantly. 
On the other hand, the loop equations technology in the gauge theory also advanced over the last decades. The correspondence between the loop space functionals and the original vector fields was better understood, and various solutions to the gauge loop equations were found. 
In particular, the momentum loop equation was developed, similar to our momentum loop used below \cite{MLDMig86, SeqQuanM95}.
Recently, some numerical methods were found to solve loop equations beyond perturbation theory \cite{LoopEqBootstrap, KazakovLooqBootstrap}.
The loop dynamics was extended to quantum gravity, where it was used to study nonperturbative phenomena \cite{AshtekarVar, RovelliSmolin}.

All these old and new developments made loop equations a major nonperturbative approach to gauge field theory.
So, it is time to revive the hibernating theory of the loop equations in turbulence, where these equations are much simpler.
The latest DNS \cite{S19, S21, VortexGasCirculation20, QuantumCirculation21} with Reynolds numbers of tens of thousands revealed and quantified violations of the K41 scaling laws. These numerical experiments are in agreement with so-called multifractal scaling laws \cite{FP85}. 

Theoretically, we studied the loop equation in the confinement region (large circulation over large loop $C$) and justified the Area law suggested in '93 on heuristic arguments.
This law says that the tails of velocity circulation PDF in the confinement region are functions of the minimal area inside this loop.
 It was verified in DNS four years ago \cite{S19}, which triggered the further development of the geometric theory of turbulence\cite{ VortexGasCirculation20, QuantumCirculation21, M23PR}.
 In particular, the Area law was justified for flat and quadratic minimal surfaces, and an exact scaling law in confinement region $\Gamma \propto \sqrt{Area}$ was derived  \cite{M23PR}. The area law was verified with better precision in \cite{S21}.

\subsection{Sparse vorticity picture}
We know (or at least assume) that the vortex structures in this extreme turbulent flow collapse into thin clusters in physical space. 
Snapshots of vorticity in numerical simulations  \cite{VTubes, Tubes19} show a collection of tube-like structures relatively sparsely distributed in space.
The large vorticity domains and the large strain domain lead to anomalous dissipation. These domains dominate the enstrophy integral so that the viscosity factor in front of this integral is compensated, leading to a finite dissipation rate. It was observed years ago and studied in the DNS\cite{SreeniDissipation}.
An excellent recent DNS \cite{S19, S21} studied statistics of vorticity structures in isotropic turbulence with a high Reynolds number. Their distribution of velocity circulation appeared compatible with circular vortex structures (which we now call Kelvinons).

The authors of \cite{S19, S21} confirmed the area law and compared it with the tensor area law, which would correspond to a constant uniform vortex, irrelevant to the turbulence. These two laws are indistinguishable for simple loops like a square, and they both are solutions to the loop equation but are very different for twisted or non-planar loops.
Their data supported the area law, inspiring our search for the relevant vorticity structures behind it.
Some recent works also modeled sparse vortex structures in classical \cite{VortexGasCirculation20} and quantum \cite{QuantumCirculation21} turbulence.

We know such singular structures in Euler dynamics: vortex surfaces and lines. Vorticity collapses into a thin boundary layer around the surface, 
 or the core surrounding the vortex line, moving in a self-generated velocity field.
The vortex surface motion is known to be unstable against the \KH{} instabilities, which undermines the whole idea of random vortex surfaces.
However, the exact solutions of the \NS{}  equations discovered in the previous century by Burgers\cite{BURGERS1948, BurgersVortex} and Townsend \cite{TW51} show stable planar sheets with Gaussian profile of the vorticity in the normal direction, peaked at the plane. 
Thus, the viscosity effects in certain cases suppress the \KH{} instabilities, leading to stable, steady vortex surfaces.
The same applies to the Burgers vortex, corresponding to a cylindrical core, regularizing a singular vortex line.
The Burgers cylindrical vortex has a constant strain, which is not the most general; there is only one independent eigenvalue instead of two.
The general asymmetric solution approaches the symmetric Burgers vortex with the potential background flow as it was subsequently found in\cite{BMO84} in the turbulent limit $|\Gamma| \gg \nu$ with asymmetric strain.

Some singular (weak ) solutions of Euler equations with nontrivial topology are described by the so-called \CL{} field, defined modulo gauge transformations.
These transformations manifestly preserve vorticity and, therefore, velocity.
These variables and their ambiguity were known for centuries \cite{Lamb45}. They were first utilized within hydrodynamics in the pioneering work of Khalatnikov \cite{khalat52} and subsequent publications of Kuznetzov and Mikhailov \cite{KM80}, and Levich \cite{L81} in the early 80-ties.
The modern mathematical formulation using \SYM{} was initiated in \cite{M83}. Yakhot and Zakharov \cite{YZ93} derived the K41 spectrum in weak turbulence using some approximation to the kinetic equations in \CL{} variables. 
The topological singular Euler solutions (vortex sheets and lines)  are similar to those studied theoretically and experimentally in superfluids \cite{Volovik1987, VollhardtWolfle, Volovik2003, Blaha1976, VolovikMineev1976, Volovik2000, KM80}. 

In a recent paper, \cite{M22}, we studied the topological circular vortex in a classical turbulent flow. In particular, we found an effective Euler Hamiltonian, with extra anomalous terms coming from the line singularity resolved as a Burgers vortex in the local tangent direction. These anomalous terms have logarithmic dependence on the Reynolds number (circulation divided by viscosity). We summed up the leading log terms, which resulted in extra powers of the logarithm of the scale, modifying the K41 scaling laws.
These exciting developments explain and quantitatively describe many interesting phenomena but do not provide a complete microscopic theory covering the full inertial range of turbulence without simplifying assumptions of the sparsity of vortex structures.
In the present work, we develop the theory free of these assumptions and approximations by solving the loop equations for decaying turbulence. 
Our solution is irregular (local vorticity and local enstrophy limit do not exist), thus resembling the Tao conjecture \cite{Tao2016}.

There is an important difference, however.
We are not studying singular solutions of the \NS{} equations; rather, we are solving the Hopf equations for a generating functional for the distribution of velocity circulation. We are looking for a statistical ensemble of \NS{} solutions, and we arrive at the probability distribution. This distribution is singular in that expectation values of powers of local vorticity diverge in a local limit.
At the same time, the correlation functions have singularities at coinciding points.
The correlation functions of vorticity at separated points are studied in the forthcoming paper by numerical simulation of our theory. These correlations have power singularities at coinciding points, showing new fractal dimensions.

\section{Loop equation}
\subsection{Wilson loop as reduced Hopf functional}
We introduced the loop equation in the Lecture Series at Cargese and Chernogolovka Summer Schools (see references in \cite{M23PR}). 
Here is a summary for the new generation.
We write the \NS{} equation as follows (with Einstein's notation of summation over repeated Greek indexes)
\begin{eqnarray}
    &&\partial_t \val = \nu \partial_{\beta} \omega_{\beta \alpha} - v_{\beta}\omega_{\beta \alpha} - \dal \left( p + \frac{\vbe^2}{2} \right);\\
    && \dal \val =0;
\end{eqnarray}

The Wilson loop average for the turbulence is defined as
\begin{eqnarray}\label{WilsonLoop}
    \Psi[\gamma,C] = \VEV{\exp{ \frac{\I \gamma}{\nu}\oint_C \val d \ral}}
\end{eqnarray}

We added a dimensionless factor $\gamma$ in the exponential compared to some previous definitions as an extra parameter of the Wilson loop. Without loss of generality, we shall assume that $\gamma >0$. The negative $\gamma$ corresponds to a complex conjugation of the Wilson loop.
In Abelian gauge theory, this parameter would be the continuous electric charge. 
The statistical averaging $\VEV{\dots}$ corresponds to initial randomized data, to be specified later.

Comparing the Wilson loop with the Hopf functional (see \cite{Hopf19} for recent review)
\begin{eqnarray}
    H[\vec J] = \VEV{\exp{\int_{\vec r\in \mathbb{R}_d} \vec J(\vec r) \cdot \vec v(\vec r)}}
\end{eqnarray}
we observe that the Wilson loop is a particular case of the Hopf functional with the source $\vec J(\vec r)$ concentrated on a fixed loop in space
\begin{eqnarray}
    \vec J(\vec r) = \frac{\imath \gamma}{\nu} \oint d \vec C(\theta) \delta\left(\vec r - \vec C(\theta)\right)
\end{eqnarray}

Three comments are necessary here. 
\begin{itemize}
    \item The Wilson loop is a dimensional reduction of the Hopf functional. Instead of a source $\vec J(\vec r)$ mapping $\mathbb{R}_d \mapsto \mathbb{R}_d$ we have a source $\vec C(\theta)$ mapping a circle to physical space $\mathbb{S}_1 \mapsto  \mathbb{R}_d $. This mapping makes the loop equation a one-dimensional problem with physical space becoming a target space, like in string theory. The functional differential equations in more than one dimension are never solved except for the linear (or linearizable) ones. One-dimensional problems, even nonlinear ones, are sometimes solvable, and the loop equation is no exception, as we shall see. Rather than being an abstract notion, the Hopf equation becomes an analytical tool after this projection to one dimension.
    \item We have lost most of the information about vector field $\vec v(\vec r)$ by reducing the Hopf source $\vec J(\vec r)$ to a one-dimensional subset $\vec C$ of the physical space. This information would be irrecoverable if we only consider the set of smooth closed loops in space. However, with the loop functional defined for arbitrary non-smooth loop $C$, the information about velocity correlation functions and other statistical observables can be recovered \cite{ M23PR} as follows. The small variation of the loop corresponding to attaching a closed loop at some point (which necessarily makes the loop non-smooth) brings down vorticity at this point. It allows to computation of the vorticity correlation functions. After that, velocity correlations can be recovered from incompressibility (the \BS{} integral in coordinate space or orthogonal projection in Fourier space).
    \item The imaginary factor $\imath \gamma$ in the source is necessary for the mathematical existence of this functional. As the DNS shows (\cite{S19, S21, KS21} ), and the Kelvinon theory supports\cite{M22}, the PDF of circulation decays only exponentially. The Fourier transform of such PDF exists but in the case of real instead of imaginary factor $\imath \gamma = \gamma_R$, the integral over the circulation distribution would diverge either on the positive or the negative circulation, depending upon the sign of $\gamma_R$; it should then be defined as an analytic continuation (Laplace transform).
\end{itemize}

With imaginary factor $\imath \gamma$, this Wilson loop becomes the wave function of the quantum mechanical system in loop space.
When treated as a function of time and a functional of the periodic function $C: \ral = C_\alpha(\theta); \;\theta \in (0,2 \pi)$, the Wilson loop satisfies the following functional Schrödinger equation (with operators $\hat \omega, \hat v$ defined below for reader's convenience)
\begin{subequations}
\begin{eqnarray}\label{OldEq}
 &&  \I \nu\partial_t \Psi = {\cal H}_C\Psi ;\\
&&{\cal H}_C =   {\cal H}^{(1)}_C + {\cal H}^{(2)}_C\\
&& {\cal H}^{(1)}_C =
    \nu \gamma\oint_{C} d \ral \dbe \hat \omega_{\alpha\beta}(r);\\
&&{\cal H}^{(2)}_C =
\gamma \oint_{C}d \ral 
\hat \omega_{\alpha\beta}(r) \hat v_{\beta }(r) ;\\
&& \hat \omega_{\alpha\beta} \equiv - \I \frac{\nu}{\gamma} \ff{\sigma_{\alpha\beta}}\\
&& \hat v_{\beta }(r)  =
\frac{1}{\partial_{\mu}^{2}}\partial_{\alpha} \hat \omega_{\beta \alpha}(r)
\end{eqnarray}
\end{subequations}

The area derivative $\ff{\sigma_{\alpha\beta}}$ is related to the variation of the functional when the little closed loop $\delta C$ is added
\begin{eqnarray}
\label{Variation}
    &&\Sigma_{\alpha\beta}(\delta C)\frac{\delta F[C]}{\delta \sigma_{\alpha\beta }(r)} = F[C+ \delta C] - F[C];\\
    \label{SigmaAlphaBeta}
    && \Sigma_{\alpha\beta}(\delta C) = \frac{1}{2} \oint_{\delta C} \ral d \rbe 
\end{eqnarray}

In the review, \cite{ M23PR},  we present the explicit limiting procedure needed to define these functional derivatives in terms of finite variations of the loop while keeping it closed.
All the operators $\partial_\mu, \hat \omega_{\alpha\beta} , \hat v_\alpha$ are expressed in terms of the spike operator
\begin{equation}
D_{\alpha}(\theta,\epsilon) = \int_{-\epsilon}^{+\epsilon}d \xi
\left(1-\frac{|\xi|}{\epsilon} \right)
    \frac{\delta}{\delta C_{\alpha}(\theta+ \xi)}
\end{equation}

The area derivative operator can be regularized as
\begin{eqnarray}\label{OM}
&&\Omega_{\alpha\beta}(\theta,\epsilon) =- \I \frac{\nu}{\gamma}\frac{\delta}{\delta C'_\alpha(\theta)}\int_{-\epsilon}^\epsilon d \xi  \frac{\delta}{\delta C_\beta(\theta+\xi)}
- \{\alpha \leftrightarrow\beta\};
\end{eqnarray}
and velocity operator (with $\delta , \epsilon \to 0^+$)
\begin{equation}\label{VOM}
V_{\alpha}(\theta,\epsilon,\delta) = \frac{1}{D_{\mu}^2(\theta,\epsilon)}
D_{\beta}(\theta,\epsilon) \Omega_{\beta \alpha}(\theta,\delta);
\end{equation}

In addition to the loop equation, every valid loop functional $F[C]$ must satisfy the Bianchi constraint \cite{MMEq79, Mig83}
\begin{eqnarray}
   \dal \frac{\delta F[C]}{\delta \sigma_{\beta\gamma }(r)} + \textbf{cyclic}=0
\end{eqnarray}

In three dimensions, it follows from identity $\vec \nabla \cdot \vec \omega =0$; in general dimension $d>3$, the dual vorticity $\tilde \omega$ is an antisymmetric tensor with $d-2$ components. The divergence of this tensor equals zero identically.
However, for the loop functional, this restriction is not an identity; it reflects that this functional is a function of a circulation of some vector field, averaged by some set of parameters.
This constraint was analyzed in \cite{M23PR} in the confinement region of large loops, where it was used to predict the Area law. The area derivative of the area of some smooth surface inside a large loop reduces to a local normal vector. The Bianchi constraint is equivalent to the Plateau equation for a minimal surface (mean external curvature equals zero).

In the \NS{} equation, we did NOT add artificial random forces, choosing instead to randomize the initial data for the velocity field.
These ad hoc random forces would lead to the potential term \cite{M23PR} in the loop Hamiltonian ${\cal H}_C $, breaking the translational symmetry in the loop space needed for the dimensional reduction we study below. What is worse, the random forces pollute the genuine turbulent dynamics with long-range artifacts, violating the universality of turbulence as a critical phenomenon. 
With random initial data instead of time-dependent delta-correlated random forcing, we no longer describe the steady state (i.e., statistical equilibrium) but decaying turbulence, which is also an interesting process, manifesting the same critical phenomena. 
The energy is pumped in before the initial moment, at the interval $-t_0 <t <0$ and slowly dissipates over time, provided the viscosity is small enough, corresponding to the large Reynolds number we are studying.

\subsection{Plane wave in loop space}

Our crucial observation in '93 was that the right side of the Loop equation, without random forcing, dramatically simplifies in functional Fourier space.
The dynamics of the loop field can be reproduced in an Ansatz
\begin{equation}\label{reduced}
  \Psi[\gamma,C] = \VEV{\exp{
	   \frac{\I \gamma}{\nu}\oint d C_{\alpha}(\theta) P_{\alpha}(\theta)}} 
\end{equation}

The difference with the original definition of $\Psi[\gamma,C]$ is that our new
vector function $ \vec P(\theta) $ depends directly on $ \theta $ rather
then through the vector field $ \vec  v(\vec r) $ in  $\mathbb{R}_d$ projected at $ \vec r =
\vec C(\theta) $. 
This transformation manifests the dimensional reduction $ d \Ra 1 $ we mentioned above. 

From the point of view of the quantum analogy, this Anzatz is a plane wave in the Loop space, solving the Schrödinger equation in the absence of forces (potential terms in the Hamiltonian, breaking the translation invariance in the loop space).
At the technical level, with the Anzatz \eqref{reduced} would pass through the loop equation provided the momentum loop $\vec P(\theta)$ does not depend on the space loop $\vec C(\theta)$
\begin{eqnarray}
    \fbyf{P_\alpha(\theta)}{C_\beta(\theta')} = 0.
\end{eqnarray}

In this case, the variations of \eqref{reduced} bring down the product of momentum
\begin{eqnarray}
    \ff{C_{\beta_1}(\theta_1)}\dots \ff{C_{\beta_n}(\theta_n)} \exp{
	   \frac{\I \gamma}{\nu}\oint d C_{\alpha}(\theta) P_{\alpha}(\theta)} \propto P'_{\beta_1}(\theta_1)\dots P'_{\beta_n}(\theta_n)
\end{eqnarray}
or, in a more general case, an arbitrary functional of the loop derivative reduces to that of the momentum loop
\begin{equation}
  \Phi\left[\frac{\delta}{\delta C_{\alpha}(\theta)}\right] \exp{
	   \frac{\I \gamma}{\nu}\oint d C_{\alpha}(\theta) P_{\alpha}(\theta)}=
	 \Phi\left[ -\frac{\I \gamma}{\nu} P'_{\alpha}(\theta)\right] \exp{
	   \frac{\I \gamma}{\nu}\oint d C_{\alpha}(\theta) P_{\alpha}(\theta)}
\end{equation}

This reduction will make the loop equation an algebraic rather than a functional differential equation. Instead of functional derivatives, we shall have ordinary functions. 
This equation appears solvable by an extra piece of luck.

\subsection{Random global vorticity}\label{RandomRot}

Let us study the momentum loop dynamics on a simple example from original papers \cite{ M23PR}.
The simplest meaningful distribution of the velocity field is the Gaussian one, with energy concentrated in the macroscopic
motions. The corresponding loop field reads (we set $\gamma =1$ for simplicity in this section)
\begin{eqnarray}
  &&\Psi_0[C] =\exp{
	 -\frac{1}{2 \nu^2}\IINT{C} d \vec C(\theta) \cdot d \vec C(\theta') f\left(\vec C(\theta)-\vec C(\theta')\right)
	}
\end{eqnarray}
where $ f(\vec r) $ is the velocity correlation function
\begin{equation}
  \left \langle v_{\alpha}(r) v_{\beta}(r') \right \rangle =
\left(\delta_{\alpha \beta}- \partial_{\alpha} \partial_{\beta}
\partial_{\mu}^{-2} \right) f(r-r')
\end{equation}

The potential part drops out in the closed loop integral.
The correlation function varies at the macroscopic scale, which means that one could expand it in the Taylor series
\begin{equation}
  f(r-r') \rightarrow f_0 - f_1 (r-r')^2 + \dots \label{Taylor}
\end{equation}

The first term $ f_0 $ is proportional to initial energy density,
\begin{equation}
  \frac{1}{2} \left \langle v_{\alpha}^2 \right \rangle =\frac{d-1}{2}
f_0
\end{equation}
and the second one is proportional to initial energy dissipation
rate ${\mathcal E}_{0}$
\begin{equation}
f_1 = \frac{{\mathcal E}_{0}}{2 d(d-1) \nu}
\end{equation}
where $ d=3 $ is the dimension of space.
The constant term in (\ref{Taylor}) as well as $ r^2 + r'^2 $ terms
drop from the closed
loop integral, so we are left with the cross-term $ r r' $, which reduces to a full square
\begin{eqnarray}\label{InitPsi}
  &&\Psi_0[C] \to \exp{- \frac{f_1}{\nu^2}\left(\oint dC_{\alpha}(\theta) C_{\beta}(\theta)\right)^2}
\end{eqnarray}

This distribution is almost Gaussian: it reduces to Gaussian one by
extra integration
\begin{eqnarray}
  &&\Psi_0[C] \rightarrow \const{}\int (d \phi) \exp{ -\phi_{\alpha \beta}^2}\nonumber\\
  &&\exp{
	 2\imath \frac{\sqrt{f_1}}{\nu}
	 \phi_{\mu\nu} \oint dC_{\mu}(\theta) C_{\nu}(\theta)}
\end{eqnarray}

The integration here involves all  $ \frac{d(d-1)}{2} =3 $ independent $ \alpha < \beta $
components of the antisymmetric tensor $ \phi_{\alpha \beta} $.
Note that this is ordinary integration, not the
functional one. 

The physical meaning of this $ \phi $ is the random
uniform vorticity $ \hat \omega = \sqrt{f_1} \hat \phi$ at the initial moment.
However, as we see it now, this initial data represents a spurious fixed point unrelated to the turbulence problem.
It was discussed in our review paper \cite{M23PR}. The uniform global rotation represents a fixed point of the \NS{} equation for arbitrary uniform vorticity tensor.
Gaussian integration by $\phi$ keeps it as a fixed point of the Loop equation.
The right side of the \NS{} equation vanishes at this special initial data so that the exact solution of the loop equation with this initial data equals its initial value \eqref{InitPsi}. 
Naturally, the time derivative of the momentum loop with the corresponding initial data will vanish as well.

It is instructive to look at the momentum trajectory $P_\alpha(\theta)$ for this fixed point.
The functional Fourier transform \cite{M23PR} leads to the following simple result
for the initial values of $P_\alpha(\theta)$.
In terms of Fourier harmonics, this initial data read
\begin{eqnarray}
  &&  P_\alpha(\theta)= \sum_{\text{odd }n=1}^\infty P_{\alpha,n} \exp{\I n \theta} + \bar{P}_{\alpha,n} \exp{-\I n \theta};\\
  && P_{\alpha,n} = \mathcal N(0,1) ;\\
  && \bar{P}_{\alpha,n} =\frac{4 \sqrt{f_1}}{n \nu} \phi_{\alpha \beta}P_{\beta,n} ;\\
  && \phi_{ \alpha\beta} = - \phi_{\beta\alpha};\\
  && \phi_{\alpha\beta} = \mathcal N(0,1) \forall \alpha < \beta;
\end{eqnarray}

The constant part $ P_{\alpha,0} $ of $ P_{\alpha}(\theta) $  is not defined, but it drops from equations by translational
invariance.
Note that this initial data is not real, as $\bar{P}_{\alpha,n}  \neq P^\star_{\alpha,n}$ .
Positive and negative harmonics are real but unequal, leading to a complex Fourier transform.
At fixed tensor $\phi$ the correlations are
\begin{eqnarray}
  &&\VEV{P_{\alpha,n} P_{\beta,m}}_{t=0}
= \frac{4 \sqrt{f_1}}{m \nu} \delta_{n m} \phi_{\alpha \beta};\\
\label{Pcorr}
&&\VEV{ P_{\alpha}(\theta) P_{\beta}(\theta')}_{t=0} =
2\imath \frac{\sqrt{f_1}}{\nu}\phi_{\alpha \beta} \sign(\theta'-\theta);
\end{eqnarray}

 This correlation function immediately leads to the uniform expectation value of the vorticity
 \begin{equation}
     \VEV{ P_{\alpha}(\theta) \Delta P_{\beta}(\theta)} = 4 \imath\frac{\sqrt{f_1}}{\nu}\phi_{\alpha \beta} ;\; \forall \theta
 \end{equation}
 
 The uniform constant vorticity kills the linear term of the \NS{} equation in the original loop space, involving $\dal \hat \Omega_{\alpha\beta}=0$. 
The nonlinear term $\hat V_\alpha \hat  \Omega_{\alpha\beta}$ vanishes in the coordinate loop space only after integration around the loop. 
 Here are the steps involved
 \begin{eqnarray}
     &&\hat V_\beta = \frac{1}{2} \hat \Omega_{\alpha\beta} C_\beta;\\
     &&\oint \hat\Omega_{\alpha\beta} C_\beta  \hat \Omega_{\beta\gamma} d C_\alpha \propto \hat\Omega_{\alpha\beta}  \hat \Omega_{\beta\gamma} \Sigma_{\alpha\beta}(C);
 \end{eqnarray}
 
 The tensor area $\Sigma$ was defined in \eqref{SigmaAlphaBeta}. It is an antisymmetric tensor; therefore its trace with a symmetric tensor $\hat\Omega_{\alpha\beta}  \hat \Omega_{\beta\gamma}$ vanishes.

 This calculation demonstrates how an arbitrary uniform vorticity tensor satisfies the loop equation in coordinate loop space.
 We expect the turbulent solution of the loop equation to be more general, with the local vorticity tensor at the loop becoming a random variable with a non-Gaussian distribution for every point on the loop.

\subsection{Reduced Loop Equation}
Let us use the \NS{} loop equation to derive a dynamical equation for the momentum loop $\vec P(t,\theta)$.
The reduced dynamics  must be equivalent to the Navier-Stokes
dynamics of the original field. With the loop calculus developed above, we
have all the necessary tools to ensure this equivalence.
Let us stress an important point: the function $\vec P(\theta, t)$ is independent of the loop $C$. As we shall see later, it is a random variable with a universal distribution in functional space.

This independence removes objection to the Kelvinon theory \cite{M22} and any other \NS{} stationary solutions with a singularity at fixed loop $C$ in space. Such solutions do not satisfy the loop equations because the functional derivatives would also act on the velocity field by varying its boundary conditions at the loop $C$. These variations would lead to extra terms to the \NS{} equation for the circulation of the velocity field.

The loop equation for $P(\theta)$ as a function of $\theta$ and also a function of time, reads:
\begin{equation}
  \partial_t  P_{\alpha} = \left(\nu D_{\beta}  -V_{\beta}\right)\Omega_{\beta \alpha}
 \label{Pdot}
\end{equation}
where the operators $ V , D, \Omega $ should
be regarded as ordinary numbers, with the following definitions.
The spike derivative $D$ in the above equation
\begin{eqnarray}\label{SpikeDer}
  &&D_{\alpha}(\theta,\epsilon) =-\frac{\I \gamma}{\nu}
 \int_{-1}^{1}d \mu
	 \mbox{ sgn}(\mu)
	 P_{\alpha} \left(\theta + \epsilon \mu \right)
	 \label{DP}
\end{eqnarray}

The vorticity \eqref{OM} and velocity \eqref{VOM} also become singular functionals of the trajectory $P(\theta)$.

The first observation about this equation is that the viscosity factor cancels after the substitution  \eqref{SpikeDer}.
As we shall see, the viscosity enters initial data so that at any finite time $t$, the solution for $P$ still depends on viscosity.
Another observation is that the spike derivative $D(\theta, \eps)$ turns to the discontinuity  $\Delta P(\theta) = P(\theta^+) - P(\theta^-)$ in the limit $\eps \to 0^+$
\begin{eqnarray}
    D(\theta,0^+) = -\frac{\I \gamma}{\nu} \Delta P(\theta)
\end{eqnarray}

The relation of the operators in the QCD loop equation to the discontinuities of the momentum loop was noticed, justified, and investigated in \cite{SeqQuanM95, Mig98Hidden}.
The momentum loop in QCD could be piecewise constant with an arbitrary number of such discontinuities.
In the \NS{} theory, the discontinuities must be present at every point on a parametric circle $\mathbb{S}_1$.

We find the local limit for vorticity
\begin{eqnarray}\label{Omega}
    &&\Omega_{\alpha\beta}(\theta,0^+) = \frac{-\I \gamma}{\nu} P_{\alpha\beta}(\theta);\\
    &&P_{\alpha\beta}(\theta) = \Delta P_\alpha(\theta)P_\beta(\theta) - \{\alpha \leftrightarrow\beta\};\\
    && P_\alpha(\theta) \equiv \frac{ P_\alpha(\theta^+) + P_\alpha(\theta^-)}{2}
\end{eqnarray}
and velocity (skipping the common argument $\theta$ )
\begin{eqnarray}
&&V_{\alpha} = \frac{\Delta P_\beta}{\Delta P_\mu^2} P_{\beta\alpha}= P_\alpha - \frac{\Delta P_\alpha \Delta P_\beta P_\beta}{\Delta P^2} 
\end{eqnarray}

The Bianchi constraint is identically satisfied as it should
\begin{eqnarray}
     \Delta P_\alpha \left(\Delta P_\beta  P_\gamma - \{\beta \leftrightarrow\gamma\}\right)  + cyclic=0
\end{eqnarray}

We arrive at a singular loop equation for $P_\alpha(\theta)$
\begin{eqnarray}\label{PloopEq}
  && \nu\partial_t \vec P =  - \gamma^2(\Delta \vec P)^2 \vec P  + \nonumber\\
  && \Delta \vec P \left(\gamma^2 \vec P \cdot \Delta \vec P +\I \gamma \left( \frac{(\vec P \cdot \Delta \vec P)^2}{\Delta \vec P^2}- \vec P^2\right)\right);
\end{eqnarray}

This equation is complex due to the irreversible dissipation effects in the \NS{} equation.
The viscosity dropped from the right side of this equation; it can be absorbed in units of time.
Viscosity also enters the initial data, as we shall see in the next section on the example of the random rotation.
However, the large-time asymptotic behavior of the solution would be universal, as it should be in the Turbulent flow.

We are looking for a degenerate fixed point \cite{M23PR}, a fixed manifold with some internal degrees of freedom. The spontaneous stochastization corresponds to random values of these hidden internal parameters.
Starting with different initial data, the trajectory $\vec P(\theta, t)$ would approach this fixed manifold at some arbitrary point and then keep moving around it, covering it with some probability measure.
The Turbulence problem is to find this manifold and determine this probability measure.

\subsection{Decay or fixed point}

The absolute value of loop average $\Psi[\gamma, C]$ stays below $1$ at any time, which leaves two possible scenarios for its behavior at a large time.

\begin{eqnarray}
    Decay: &&\vec P \to 0;\; \Psi[\gamma,C] \to 1; \\
    Fixed Point:  &&\vec P \to \vec P_\infty;\; \Psi[\gamma,C] \to \Psi_\infty[C];
\end{eqnarray}

The Decay scenario in the nonlinear ODE \eqref{PloopEq} corresponds to the $1/\sqrt{t}$ decrease of $\vec P$.
Omitting the common argument $\theta$, we get the following exact time-dependent solution (not just asymptotically, at $t \to +\infty$).
\begin{eqnarray}\label{decayingSolution}
   && \vec P = \sqrt{\frac{\nu}{2(t+ t_0)}} \frac{ \vec F}{\gamma};\\
   \label{FEQ}
   && \left((\Delta \vec F)^2 -1\right) \vec F =  \nonumber\\
  && \Delta \vec F \left( \vec F \cdot \Delta \vec F +\frac{\I }{\gamma}\left( \frac{(\vec F\cdot \Delta \vec F)^2}{(\Delta \vec F)^2}- \vec F^2\right)\right);
\end{eqnarray}

The fixed point would correspond to the vanishing right side of the momentum loop equation \eqref{PloopEq}. Multiplying by $(\Delta \vec P)^2$ and reducing the terms, we find a singular algebraic equation
\begin{eqnarray}
   \label{GEQ}
   &&\gamma^2 (\Delta \vec P)^2\left((\Delta \vec P)^2 \vec P - (\vec P \cdot \Delta \vec P)\Delta \vec P \right) =\nonumber\\
  &&  \I \gamma\Delta \vec P \cdot\left( (\vec P \cdot \Delta \vec P)^2- \vec P^2 (\Delta \vec P)^2\right);
\end{eqnarray}

The fixed point could mean self-sustained turbulence, which would be too good to be true, violating the second law of Thermodynamics. 
Indeed, it is easy to see that this fixed point cannot exist.
The fixed point equation \eqref{GEQ} is a linear relation between two vectors $\vec P, \Delta \vec P$ with coefficients depending on various scalar products.
The generic solution is simply
\begin{eqnarray}
    \Delta \vec P =\lambda \vec P; 
\end{eqnarray}
with the complex parameter $\lambda$ to be determined from the equation \eqref{GEQ}.
This solution is degenerate: the fixed point equation is satisfied for arbitrary complex $\lambda$.
The discontinuity vector $\Delta \vec P$ aligned with the principal value $\vec P$ corresponds to vanishing vorticity in \eqref{Omega}, leading to a trivial solution of the loop equation $\Psi[\gamma, C]=1$.
We are left with the decaying turbulence scenario \eqref{FEQ} as the only remaining physical solution.
\section{Fractal curve in complex space}\label{RWCS}
\subsection{Random walk}
One may try the solution where the discontinuity vector is proportional to the principal value.
However, in this case, such a solution does not exist. Assuming such a solution, we are led to the following contradictory equations
\begin{eqnarray}
    && \Delta \vec F \stackrel{?}{=}\lambda \vec F;\\
    && \lambda^2 \vec F^2  -1  \stackrel{?}{=} \lambda^2 \vec F^2;
\end{eqnarray}

There is, however, another solution where the vectors $\Delta \vec F, \vec F$ are not aligned.
This solution requires the following relations
\begin{subequations}\label{GapEq}
\begin{eqnarray}
    &&(\Delta \vec F)^2 =1;\\
    && (2 \vec F \cdot \Delta \vec F- \imath \gamma)^2  + \gamma^2 = 4\vec F^2 
\end{eqnarray}
\end{subequations}

A simple algebra using these two relations shows that both sides of the decay equation \eqref{FEQ} vanish: $0 \Delta \vec F = 0 \vec F$.
At the same time, vectors $\Delta \vec F$ and $ \vec F$ will not be aligned, so the vorticity remains finite.
Conversely, without these relations, the vectors $\Delta \vec F$ and $ \vec F$ will be aligned, leading to vanishing vorticity in \eqref{Omega}.

These relations are very interesting. The complex numbers reflect irreversibility, and lack of alignment leads to vorticity distributed along the loop.
\begin{eqnarray}
     &&\Omega_{\alpha\beta} =  \frac{\imath}{2 (t+ t_0)} \left(F_\alpha \Delta F_\beta - F_\beta \Delta F_\alpha\right);
\end{eqnarray}

Note that the system of equations \eqref{GapEq} is parametrically invariant (being local and independent of $\theta$).
Also, note that this complex vector $\vec F(\theta)$ is dimensionless, and this system is completely universal, up to a single dimensionless parameter $\gamma$. The viscosity dropped from this equation, and the dimension of vorticity is inverse time, which explains the factor $1/(t+ t_0)$.
These equations do not allow $\Delta \vec F =0$, so discontinuities must be present at every $\theta$. In other words, our solution cannot be smooth: it is a fractal curve rather than a piecewise smooth curve with discontinuities at several points.

We approximate this fractal curve by a polygon, with piecewise constant $\vec F(\theta)$ and $N$ gaps $\Delta \vec F(\theta_k)$ at equidistant angles $\theta_k = 2 \pi k/N$. If the limit $N \to \infty$ exists, we get the desired fractal curve $\vec F(\theta)$ with a discontinuity at every $\theta$.
One can build such a solution as a limit of the Markov process by the following method. 
Start with a complex vector $\vec F(\theta=0) = \vec F_0$.
We compute the next values $\vec F_k = \vec F\left(\frac{2 \pi k}{N}\right)$ from the discontinuity equations \eqref{GapEq}.
\begin{subequations}\label{DiscEqF}
\begin{eqnarray}
    && \left(\vec F_{k+1} - \vec F_{k} \right)^2 =1;\\
    && \left(\vec F_{k+1}^2 - \vec F_{k}^2 - \imath \gamma\right)^2 +\gamma^2= \left(\vec F_{k+1} + \vec F_{k} \right)^2 
\end{eqnarray}
\end{subequations}

\subsection{Constraints imposed on a random step}

A solution to these equations can be represented using a complex  vector $\vec q_k$ subject to two complex constraints 
\begin{subequations}\label{DiscEq}
\begin{eqnarray}
    &&\vec q_k^2=1;\\
    \label{constraintOnDot}
    &&\left( 2\vec F_k\cdot \vec q_k - \I\gamma\right)^2= 4 \vec F_k^2 +\gamma(2 \I -\gamma)
\end{eqnarray}
\end{subequations}
after which we can find the next value
\begin{eqnarray}
    && \vec F_{k +1} = \vec F_k + \vec q_k;
\end{eqnarray}

We assume $N$ steps, each with the angle shift $\Delta \theta = \frac{2 \pi}{N}$.
This recurrent sequence is a Markov process because each step only depends on the current position $\vec F_k$.
The closure requirement $\vec F_N = \vec F_0$ makes it a periodic Markov process, with the period $N$.
This requirement represents a nonlinear restriction on all the variables $\vec F_k$, which we discuss below.
With this discretization, the circulation can be expressed in terms of these steps
\begin{eqnarray}\label{FdC}
    \oint \vec F(\theta) \cdot d \vec C(\theta) = -\oint  \vec C(\theta) \cdot d \vec F(\theta) \Ra -\sum_{k=0}^{N-1} \frac{\vec C_{k+1} + \vec C_{k}}{2} \cdot \vec q_k
\end{eqnarray}

Note that the complex unit vector is \textbf{not} defined with the  Euclidean metric in  $2 d$ dimensions $ \VEV{\vec A, \vec B} = \Re \vec A \cdot \Re\vec B + \Im \vec A \cdot \Im\vec B$. 
Instead, we have a complex condition
\begin{eqnarray}
    \vec q^2 =1
\end{eqnarray}
which leads to two conditions between real and imaginary parts
\begin{eqnarray}
    &&(\Re \vec q)^2 = 1 + (\Im \vec q)^2;\\
    && \Re \vec q \cdot \Im \vec q =0;
\end{eqnarray}

In $d $ dimensions, there are $d-1$ complex parameters of the unit vector; with an extra constraint in \eqref{constraintOnDot}, there are now $d-2$ free complex parameters at every step of our iteration, plus the discrete choice of the sign of the root in the solution of the quadratic equation \eqref{constraintOnDot} for the projection $ \vec F_k \cdot \vec q_k$.

\subsection{Closure condition}
At the last step,  $ k = N-1$, we need to get a closed loop $\vec F_N = \vec F_0$. This is one more constraint on the complex vectors $\vec q_0,\dots \vec q_{N-1}$
\begin{eqnarray}
    &&\sum_0^{N-1} \vec q_k =0;
\end{eqnarray} 

We use this complex vector constraint to fix some of the remaining parameters.
Due to the closure of the space loop $\vec C(\theta)$, the global translation of the momentum loop $\vec P(\theta)$ leaves invariant the Wilson loop; this leads to certain gauge invariance (see below).
The circulation must correspond to a real number, though the Wilson loop is not real, as there is an asymmetry in the distribution of signs of circulation.
We discuss this issue in the following sections.

\subsection{Mirror pairs of solutions}
Return to the general study of the discrete loop equations \eqref{DiscEq}.
There is a trivial solution to these equations at any even $N$
\begin{eqnarray}
   && \vec f_k = \frac{(-1)^k \vec q}{2};\\
   && \vec q^2 = 1;
\end{eqnarray}

We reject this solution as unphysical: the corresponding vorticity equals zero, as all the vectors $\vec f_k $ are aligned.
Our set of  equations has certain mirror reflection symmetry
\begin{eqnarray}\label{RefSym}
    \vec F_k \leftrightarrow \vec F^*_{N-k}
\end{eqnarray}

Thus, the complex solutions come in mirror pairs $\vec F_k,\vec F^*_{N-k}$. The real solutions are only a particular case of the above trivial solution with real $\vec q$. 
Each nontrivial solution represents a periodic random walk in complex vector space $\bC^d$. The complex unit step $\vec q_k \in \bC^d$ depends on the current position $\vec F_k \in \bC^d$, or, equivalently, on the initial position $\vec F_0$  plus the sum of the preceding steps.
We are interested in the limit of infinitely many steps $N\to \infty$, corresponding to a closed fractal curve with a discontinuity at every point.

\subsection{The degenerate fixed point and its statistical meaning}

This solution's degeneracy (fewer restrictions than the number of free parameters) is a welcome feature. One would expect this from a fixed point of the Hopf equation for the probability distribution.
This degeneracy leads to stochastization of the \NS{} flow at large Reynolds numbers. The solution comes as a manifold, and the flow covers this manifold with some invariant measure.
In the best-known example, the microcanonical Gibbs distribution for Newton's mechanics covers the energy surface with a uniform measure (ergodic hypothesis, widely accepted in Physics, though still unproven in Mathematics). 
The parameters describing a point on this energy surface are not specified-- in the case of an ideal Maxwell gas, these are arbitrary velocities of particles.

Likewise, the fixed manifold, corresponding to our fractal curve, is parametrized by $N\to \infty$ sign variables, like an Ising model, plus an arbitrary global rotation matrix $\hat O\in SO(d)$ and a global parameter $\beta$, as discussed in the next section.
This rich internal random structure of our fixed manifold, combined with its rotation and translation invariance in loop space $C$, makes it an acceptable candidate for extreme isotropic turbulence.
\section{Exact analytic solution}

\subsection{Random walk on a circle}
How could a complex curve describe real circulation? This remarkable cancellation of the imaginary part of the circulation is possible if the imaginary part of $\vec P(\theta)$ does not depend on $\theta$. 
Such an imaginary term will drop after integration over closed loop $\vec C(\theta)$.

We have found a family of such solutions \cite{DecayTurb23} of our recurrent equation \eqref{DiscEqF} for arbitrary $N$
\begin{eqnarray}\label{SymSol}
    &&\vec F_k =  \frac{1}{2} \csc \left(\frac{\beta }{2}\right) \left\{\cos (\alpha_k), \sin (\alpha_k) \vec w, i \cos \left(\frac{\beta }{2}\right)\right\};
\end{eqnarray}

Here  $\vec w \in \bS^{d-3}$ is a unit vector.
The angles $\alpha_k$ must satisfy recurrent relation
\begin{eqnarray}\label{alphaEq}
    &&\alpha_{k+1} = \alpha_k + \sigma_k \beta;\\
    && \alpha_N - \alpha_0 = 0 \Mod { 2 \pi};\\
    && \sigma_k^2 =1
\end{eqnarray}

This sequence with arbitrary signs $\sigma_k=\pm1$ solves recurrent equation \eqref{DiscEqF} independently of $\gamma$.
The closure condition requires certain relations between these numbers. 
The main condition is that $\beta$ must be a rational fraction of $2 \pi$:
\begin{eqnarray}
    \beta = \frac{2 \pi p}{q} ; 0 < p < q < N;  q \neq 2;
\end{eqnarray}

The  $q=2$ case is eliminated. It corresponds to $ p =1, \beta = \pi, \vec F_k = \infty$.
Otherwise, the periodic solution for $\alpha_k$ will correspond to the following set of  $\sigma_k$
\begin{eqnarray}
   &&\sigma = \{ 1,\dots,1,-1,\dots,-1\}_{perm};
\end{eqnarray}

This array has $N_+$ positive values and $N_-$ negative values where
\begin{eqnarray}
   &&N_\pm = \floor*{\frac{N \pm  r q}{2}};\\
   && N_+ + N_- = N;\\
   && N_+ - N_- = r q - (N- r q) \Mod{2} ;
\end{eqnarray}

The symbol $perm$ stands for a random permutation of the array, which preserves its sum
\begin{eqnarray}
    \sum\sigma_k = (N_+ - N_-)
\end{eqnarray}

This sum must be a multiple of $q$ for periodicity, which leads to another restriction
\begin{eqnarray}\label{periodicity}
    (N - r q) \Mod{2} =0
\end{eqnarray} 

In other words, $ r q$ must have the same parity as $N$.

These properties lead to periodicity
\begin{eqnarray}
    \alpha_N- \alpha_0 = \beta \sum\sigma_k = 2 \pi p r
\end{eqnarray}

At fixed denominator $q$, the winding number $r$ can take the values 
\begin{eqnarray}
    &&r = \left\{-r_{max},\dots, 0,\dots, r_{max}\right\};\\
    && r_{max} = \floor{N/q}
\end{eqnarray}

The sequence with all spins flipped: $\sigma_k \Ra -\sigma_k$ also solves the loop equation. This sequence is a reflected solution we mentioned, so we include it in the statistical samples with equal probability. It corresponds to the winding number's reflection $r \Ra -r$.
The number of states with given $N, p, q, r$ is a partition of $ N_+$ positive and $N_-$ negative spins into $N$ boxes.
The probabilities are given by binomial distribution with $w = \oh$
\begin{eqnarray}
    W(N, p, q, r) = 2^{-N} \Binom{N}{N_+} = \frac{N!}{2^N N_+! N_-!}
\end{eqnarray}

The next section discusses this ensemble of rational numbers and its statistics at $N\to \infty$.
Given the rational number $\frac{p}{q}$, we can generate the sequence of angles $ \sum \pm \beta$, adding to a $2 \pi$ multiple.
The random walk step $\vec q_k  = \vec F_{k+1} - \vec F_k$ is a real unit vector in this solution
\begin{eqnarray}
    &&\vec q_k = \sigma_k\{ - \sin \delta_k, \vec w \cos \delta_k , 0\};\\
    &&\delta_k =  \alpha_k +\frac{\beta  \sigma_k}{2} 
\end{eqnarray}

The direction of this vector is not random, though; in addition to the random sign $\sigma_k$ and random unit vector $\vec w$ in $d >3$ dimensions, its direction depends on the previous position $\alpha_k$ on a circle.
So, this is a perfect example of a periodic Markov chain, with the periodicity condition analytically solved by quantizing the angular step to a rational number $\beta = \frac{2 \pi p}{q}$.
This solution corresponds to the real value of velocity circulation on each of these two solutions; however, the reflection changes this value.
Thus, the arithmetic average of two Wilson loops with two reflected solutions is reflection-symmetric, but it is still a complex number.

Our solution has a peculiar gauge invariance.
The circulation and, therefore, all observables are invariant under the shift of all $\vec F_k$ by a constant vector:
\begin{eqnarray}
    \vec F_k \Ra \vec F_k + \vec V
\end{eqnarray}

This gauge invariance follows from the closure of the loop $C$: any constant term in $\vec F(\theta)$ yields zero after integration $\oint d \vec C =0$, or summation $\sum \Delta \vec C =0$.
Using this invariance, we can drop the last component of $\vec F_k$ so that they become real vectors
\begin{eqnarray}
    &&\vec F_k \Ra  \frac{1}{2} \csc \left(\frac{\beta }{2}\right) \left\{\cos (\alpha_k), \sin (\alpha_k) \vec w, 0\right\};
\end{eqnarray}

The vorticity operator in this gauge will become a purely imaginary vector in the $z$ direction:
\begin{eqnarray}
    \vec \omega_k = \left\{0,0,\frac{\imath \sigma_k}{2} \cot\left(\frac{\beta}{2}\right)\right\}
\end{eqnarray}

As we shall see, this does not lead to complex numbers for the correlation functions of vorticity in physical space.
The correlation function of two vorticities, separated by a finite distance $\vec r$ in an "inertial interval," is finite and real after integration over the global rotation matrix. Its limit at $\vec r \to 0$ may be singular so that the anomalous dissipation may emerge.

This discrete set $N, p, q, r,\sigma_k$ describes a particular solution of the loop equation for the Wilson loop in decaying turbulence. 
Here is an important point to keep in mind. Unlike the \NS{} equation, the loop equation is linear.
Therefore, any superposition of its solutions parametrized by  $N, p, q, r,\sigma_k$ also solves the loop equation. We need to find a particular superposition that has the correct physical properties.
In particular, this superposition must describe a continuum limit of our fractal curve when $N\to \infty$. In the next two sections, we study an ensemble of such solutions,  the Euler ensemble. This ensemble corresponds to adding every solution with equal weight.
Naturally, the question arises: Why equal weight? Why not give the even $N$ more weight or exclude prime numbers?

This is our ergodic hypothesis.

Equal weight is the most symmetric option from the mathematical point of view, plus methods of numbers theory can study the properties of such an ensemble.
One argument favoring the equal weight hypothesis is that weight distribution may become irrelevant in the local limit. 
This limit, as we shall see in the rest of the paper, is determined by the statistical weight of the configurations of the ensemble. Therefore, the singular behavior of the ensemble when the mean number $\VEV{N}$ of vertices in the fractal curve goes to infinity may be universal.
In other words, the continuum fractal curve corresponding to the limit of this solution may be a unique mathematical object independent of the method used to approximate it by a polygon.

In that case, there may be an alternative way to describe this fractal curve without taking a limit of a large polygon. This description would require more sophisticated mathematical methods than those we use in this paper. 
The advantage of our polygonal approach to the fractal curve is that it is well-defined and is solvable (the ensemble averages are calculable) at every finite $N$ and can be analytically extrapolated to the ensemble with $\VEV{N} \to \infty$.
In terms of modern QFT, the continuum limit of this statistical ensemble is \textbf{dual} to the statistical theory of the velocity field.
In the same way, the discrete model of well-known dynamical triangulations is dual to the continuum theory of quantum gravity.

\subsection{The Euler ensemble }\label{EulerEns}

The discrete set of fractions $\frac{p}{q}$ with denominator $ q < N$ is well known in the number theory \cite{HardyWright}, starting with Euler.
However, our extra restriction $\sum_l\sigma_l \Mod{q} =0$ ties the set of fractions to the set $Z_2^{\otimes N}$ of $N$ Ising variables.
We will study the statistics of the corresponding ensemble, which we call the Euler ensemble, honoring great Leonard Euler.
He never thought that his $\varphi$ functions and his equation for ideal fluid would meet in theoretical physics, but good theories have a life of their own.

We distinguish between big Euler ensembles and small Euler ensembles.
The big ensemble $\mathbb{E}(N)$ assigns equal weight for each element of the large set
\begin{eqnarray}
    &&\text{variables: }  p,q,r,\left\{\sigma_1\dots\sigma_N\right\} \\
    && 0 < p  < q < N;\\
    && -N \le q r \le N;\\
    && \sigma_1,\dots \sigma_N  = \pm 1;\\
    && W_N\left(p,q,r,\left\{\sigma_1\dots\sigma_N\right\}\right) = 
    \begin{cases}
        1 & \text{if }(p,q) =1, \sum_i \sigma_i = q r\\
        0 &\text{otherwise}\\
    \end{cases}
\end{eqnarray} 

The small ensemble $\mathcal{E}(N)$ results from averaging the big ensemble over the $\sigma$ variables
\begin{eqnarray}
    &&\text{variables: }  p,q,r,\\
    && 0 < p  < q < N;\\
    && -N \le q r \le N;\\
    && w_N\left(p,q,r\right) = 
    \begin{cases}
        w_N(q, r) & \text{if } (p,q) =1\\
        0 &\text{otherwise}\\
    \end{cases}\\
    && w_N(q, r) =
    \begin{cases}
        2^{-N} \Binom{N}{(N + q r)/2} & \text{if } 2|(N- q r)\\
        0 &\text{otherwise}\\
    \end{cases}
\end{eqnarray} 

The binomial coefficients count the number of states with $\sum_i \sigma_i = q r$ among all $2^N$ states of the set of $\sigma_i=\pm 1, i =1,\dots N$.
We divided the statistical weights (the number of allowed configurations) by the total number $2^N$ of spin configurations. This normalization makes $ w_N\left(q,r\right)$  the probability of finding the values $q,r$ in the big Euler ensemble with random $\sigma_i$.

Let us count all fractions with denominator $ q < N$ and proper parity, same as $N$. All the integers between $2$ and $N$ are allowed for $q$, and each such number would enter with the weight $ \sum_r w_N(q,r)$.
At given $q < N$ the allowed numbers of $p$ are all integers $ 0 < p < q$ such that $\mathbf{gcd}(p,q) =1$. The famous Euler's totient $\varphi(q)$ \cite{HardyWright} counts such numbers.
\begin{eqnarray}\label{discretePQ}
   && Prob[q < Q] = \frac{Z(Q, N)}{Z(N, N)}\\
   && Z(Q, N) = \sum_{2<q<Q}\varphi(q)\sum_r w_N(q,r);\\
   && \varphi(q) = \sum_{\substack{p=1 \\ (p,q)=1}}^{q-1} 1
\end{eqnarray}

Thus, we can relate the big Euler ensemble average of some function to the small ensemble average as follows
\begin{eqnarray}\label{EulerAverage}
&& \VEV{F(\dots)}_{\mathbb{E}(N)} = \VEV{\VEV{F(\dots)}_\sigma}_{\mathcal{E}(N)} ;\\
 && \VEV{F\left(p,q,r,\left\{\sigma_1\dots\sigma_N\right\}\right)}_\sigma \equiv 2^{-N} \sum_{\substack{\sigma_i=\pm 1\\ \sum_i\sigma_i= q r}}F\left(p,q,r,\left\{\sigma_1\dots\sigma_N\right\}\right);\\
&& \VEV{F(p,q,r,N)}_{\mathcal{E}(N)} =\frac{\displaystyle\sum_{2<q<N}\sum_{\substack{p=1\\(p,q)=1}}^{q-1}\sum_{\substack{r\\2|(N-q r)}} F(p,q,r,N)}{Z(N,N)};
\end{eqnarray}

The advantage of this representation is that we can first average by the $\sigma$ variables, which is a rather simple arithmetic mean.
After that, we have to average over the small Euler ensemble, which involves only three variables. Resulting triple sums are calculable numerically with \Mathematica and also can be studied in the local limit $N\to \infty$ by the methods of number theory.

The odd and even ensembles have different asymptotic behavior with $N$.
Here are the allowed parity of $q,r$ for odd/even $N$
\begin{eqnarray} 
    \begin{cases}
        \text{odd }r, \text{odd }q  & \text{odd } N\\
       \text{integer } r \text{, even } q &\text{even } N\\
       \text{even } r  \text{, odd } q &\text{even } N\\
    \end{cases}
\end{eqnarray}

\begin{figure}[h]

\begin{subfigure}{0.5\textwidth}
\includegraphics[width=0.9\linewidth, height=6cm]{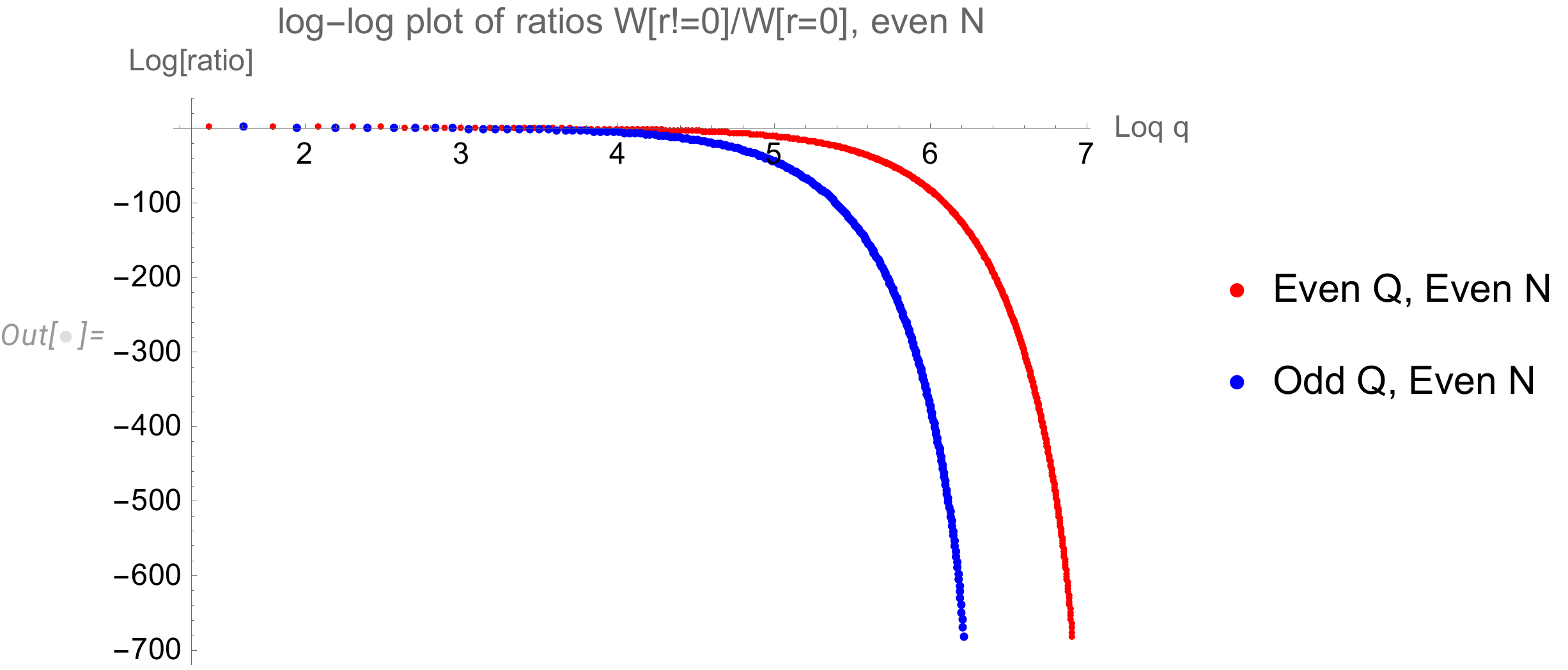} 
\caption{Even $N$}

\end{subfigure}
\begin{subfigure}{0.5\textwidth}
\includegraphics[width=0.9\linewidth, height=6cm]{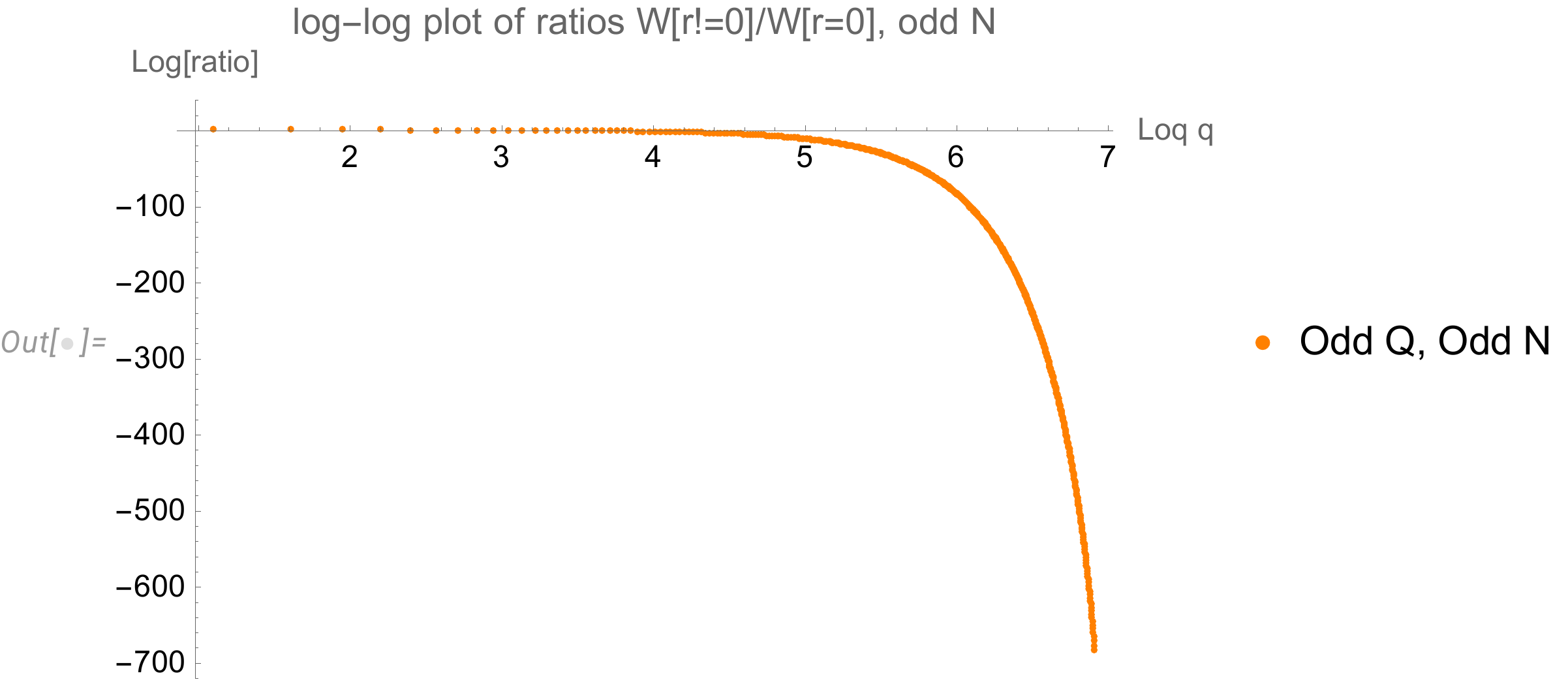}
\caption{Odd $N$}
\end{subfigure}
\caption{Log-log plots of four ratios $R_N(q) $ for even/odd $2 <q < N$, even/odd $N= 1000, 1001$.
The larger $N$ led to astronomically small ratios, so we did not use them.}
\label{fig::FourRatios}
\end{figure}

The ratios
\begin{eqnarray}
    R_N(q) = \frac{\sum_{ r\neq 0} w_N(q,r)}{ w_N(q,0)}
\end{eqnarray}
were computed for $ N = 1000, 1001$  (see Fig.\ref{fig::FourRatios}).
These ratios are finite, up to some value of $q$, after which they quickly go to zero (faster than exponentially).
This fast decrease suppresses these $r\neq 0$ terms in the sum over $q$, which otherwise diverges at the upper limit and grows as $ N^2 $.
\pct{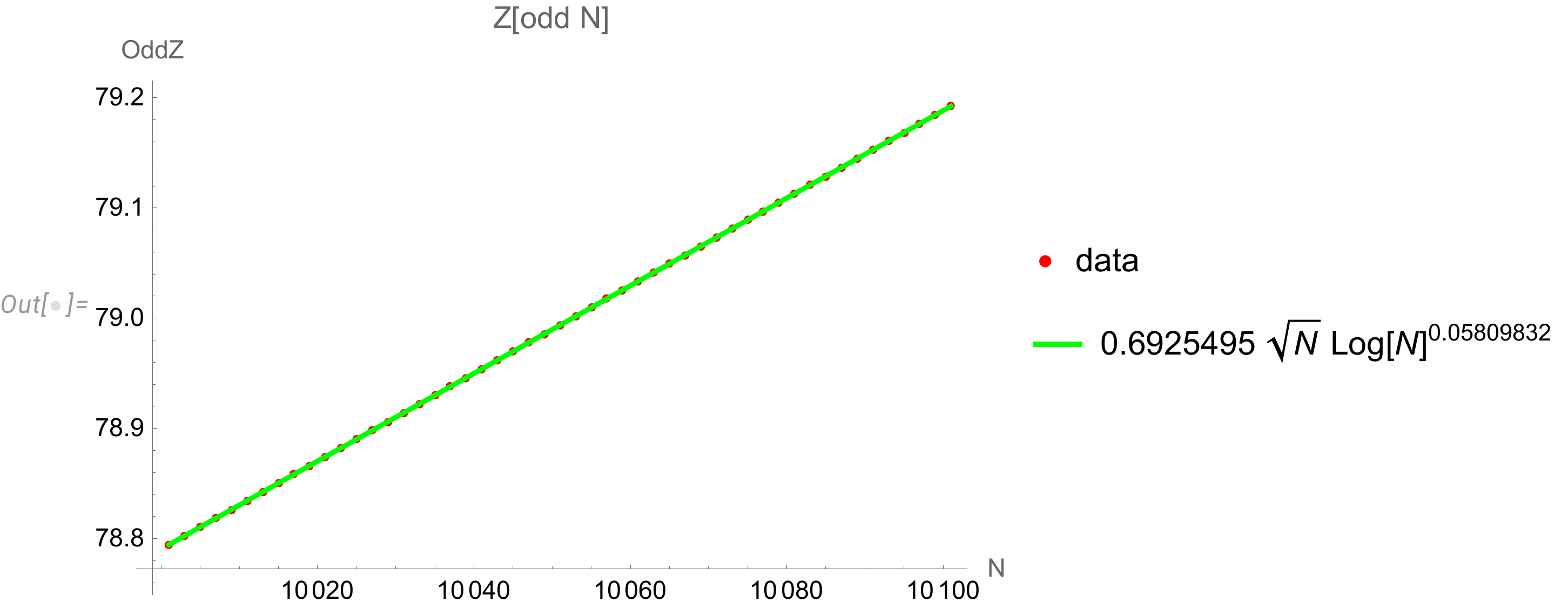}{Partition function $Z(N,N)$ for odd $N$, fitted as $ a \sqrt{N} \log(N)^b$}
At large $N$ the $r=0$ weight tends to the $q$- independent limit
\begin{eqnarray}
    w_N(q,0) = 2^{-N} \Binom{N}{\frac{N}{2}} \to \sqrt{\frac{2}{\pi N }}
\end{eqnarray}
which does not restrict the sum of Euler totients $\sum_q \varphi(q)$.
Therefore, for even $N$
\begin{eqnarray}
    Z(N,N)_{\text{even }N} \to \sqrt{\frac{2}{\pi N}}\sum_{2 < q < N} \varphi(q) \to \frac{3 \sqrt{2} N^{3/2}}{\pi ^{5/2}}
\end{eqnarray}

For odd $N$, the leading term is missing, so we have to sum the terms with $r\neq 0$. 
\begin{eqnarray}
    Z(N,N)_{\text{odd }N} =\sum_{2 < q < N} \varphi(q) \sum_{r>0} 2 w_N(q,r)
\end{eqnarray}

This term converges at $q \ll N$ where $\sum_{r>0} 2 w_N(q,r)$  is not exponentially small. The asymptotic formulas for summators of the Euler totients do not apply here, so we can compute this sum numerically and extrapolate to $N \to \infty$.
We computed it numerically in \Mathematica and fitted it to $\sqrt{N}$ times a power of $\log N$ (see Fig.\ref{fig::ZFitOddN}).
\begin{eqnarray}
    Z(N,N)_{\text{odd }N} \approx 1.2072 \sqrt{N} \log ^{0.078776}(N)
\end{eqnarray}

It is a challenge to the number theory to produce exact asymptotic behavior, replacing my fitted "law."
Recently, the number theory answered this challenge \cite{Zah23}. The result of their computation confirms the $\sqrt{N} $ factor but rejects the $\log ^{\alpha}(N)$ correction factor as an overfit due to insufficiently large $N$.
The actual pre-exponential coefficient is required for the full theory of the Euler ensemble. However, it does not contribute to the leading term of the grand canonical ensemble we use in the rest of the paper.
The results of \cite{Zah23} correspond to $Z(N,N)_{\text{odd }N} \to\sqrt{N}/\sqrt{2 \pi}$ without any factors of $\log N$. This mistake reminds us that mathematical laws must be derived theoretically rather than fitted to numerical data.

\subsection{Grand canonical ensemble}

The original Euler ensemble with fixed $N$ can be regarded as a microcanonical ensemble of statistical mechanics.
The grand canonical ensemble is more appropriate if the number of states can fluctuate, as in our case, where probabilities drastically change when $N$ is shifted by $1$.
The weights for varying numbers $N$ of degrees of freedom are multiplied by $ e^{- \mu N}$, then $N$ is treated as the rest of the variables.
The chemical potential $\mu$ will have to tend to zero in the thermodynamic limit, and the resulting singularity becomes a thermodynamic singularity corresponding to the critical phenomena.
Note that we changed the sign of $\mu $ compared to the historical definition: our $ \mu \to +0$ so that the opposite sign would be inconvenient.

The partition function and the ensemble averages in the grand canonical ensemble are
\begin{eqnarray}
   && Z(\mu)  = \sum_N Z(N,N) e^{- \mu N};\\
   && \VEV{F(p,q,r,N)}_{\mathcal{E}(\mu)} =\frac{\displaystyle\sum_N e^{- \mu N}\displaystyle\sum_{2<q<N}\sum_{\substack{p=1\\(p,q)=1}}^{q-1}\sum_{\substack{r\\2|(N-q r)}} F(p,q,r,N)}{Z(\mu) };
\end{eqnarray}

With the grand canonical ensemble, the ambiguity of the local limit disappears.
At the critical point $\mu \to 0$, the even $N$ dominate, with the following result
\begin{eqnarray}\label{Zmu}
    Z(\mu) \to \frac{1}{2}\frac{9}{2 \sqrt{2} \pi ^2 \mu ^{5/2}}
\end{eqnarray}

The extra factor of $1/2$ came from skipping all the odd values of $N$. The remaining sum over even $N$ tends to be half the asymptotic expression's integral for these even $N$.
In the rest of the paper, we shall use the grand canonical ensembles $\mathbb{E}(\mu),\mathcal{E}(\mu) $ in the thermodynamic limit $\mu \to 0$, where we find the critical phenomena. 

With the grand canonical ensemble, the ergodic hypothesis becomes a weaker restriction: any smooth weight function $W(N)$ in the distribution will cancel in expectation values in the limit $\mu \to 0$. For example, the saddle point calculation yields
\begin{eqnarray}
    \frac{\int_2^\infty d N W(N) N^\alpha F(N)\exp{- \mu N}}{\int_2^\infty d N W(N) N^\alpha\exp{- \mu N} } \to F(\alpha/\mu)
\end{eqnarray}

This phenomenon is "self-organized criticality," unlike conventional second-order phase transitions in statistical physics, which require tuning of thermodynamical potentials (temperature, pressure, chemical potential, etc.). The critical point $\mu_c =0$ does not require fine-tuning.
The self-organized criticality, or spontaneous stochastization of turbulence, is an expected property that was never proven theoretically but observed in numerical and real experiments.
In our theory, this spontaneous stochastization naturally emerges in the solution of the loop equations in the local limit.
\section{Correlation functions}
In this section and the rest of the paper, we only consider the three-dimensional space we inhabit. There are interesting mathematical problems related to decaying turbulence in higher dimensions, which we leave to future researchers. In less than three dimensions, our solutions do not exist.

\subsection{General formulas}
The simplest observable quantities we can extract from the loop functional are the vorticity correlation functions \cite{M23PR}, corresponding to the loop $C$ backtracking between two points in space $\vec r_1=0, \vec r_2=\vec r$, see Fig.\ref{fig::Backtracking}.
The vorticity operators are inserted at these two points.

The correlation function reduces to the following average over the ensemble $\mathbb{E}(\mu)$ of our random curves in complex space.
\begin{eqnarray}\label{CorrFunc}
   && \VEV{\vec\omega(\vec 0) \cdot \vec \omega(\vec r)} = \frac{\VEV{\displaystyle\sum_{0\le n<m< N}\vec \omega_m \cdot \vec \omega_n  \exp{\imath\vec \rho \cdot \left( \vec S_{n,m}- \vec S_{m,n} \right) }}_{\mathbb{E}(\mu)}}{ 4 (t+ t_0)^2} ;\\
   && \vec S_{n,m} = \frac{\sum_{k=n}^{m-1}  \vec F_k}{(m  -n) \Mod{N}};\\
   && \vec \rho = \frac{\vec r}{2\sqrt{\nu(t+ t_0)}};
\end{eqnarray}

The averaging $\VEV{\dots}$ in these formulas involves group integration $\int_{O(3)} d \hat O$ with $ \vec \rho \Ra \hat O \cdot \vec \rho$. 
\begin{eqnarray}
    \VEV{H(\vec \rho \cdot \hat \Omega \cdot \vec F)}_{O(3)} = \oh \int_{-1}^{1} d z H\left(|\vec \rho| |\vec F| z\right)
\end{eqnarray}

Let us explain the origin of summation over two positions $n,m$ of the points $\vec r_1 =0, \vec r_2 = \vec r$ on the discreet loop $\vec C(\theta)$.
There is a degree of freedom we did not specify until now, namely, the reparametrization of the momentum loop $\vec P(\theta)$.
\pct{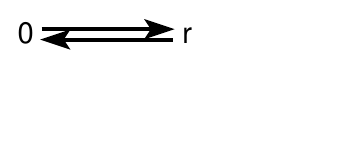}{Backtracking wires corresponding to vorticity correlation function in \eqref{CorrFunc}. With these backtracking wires, the correlation function reduces to the closed loop functional, which is represented by our solution with fractal momentum loop $\vec P(\theta)$.}
The loop equations \eqref{PloopEq} are invariant under the one-dimensional diffeomorphisms ( or reparametrizations)
\begin{eqnarray}
    &&\vec P(\theta) \Ra \vec P(f(\theta)); \\
    && f'(\theta) >0, f(2 \pi) = f(0) + 2 \pi
\end{eqnarray}

Thus, the general solution involves an arbitrary monotonous function $f(\theta)$, and averaging over the fixed manifold of the solutions of the \NS{} equations involves functional integration over all such functions.
This integration includes summation over the positions $\theta_1, \theta_2 $ of the vorticity insertion points on a curve $\vec C(\theta)$. In the continuum theory, this would be an ordered Lebesgue integration (diffeomorphisms preserve the ordering of points on a curve)
\begin{eqnarray}
    \iint_{-\pi}^\pi  \Theta(\theta_2 - \theta_1) \left(\vec P(\theta_1)\times d \vec P(\theta_1)\right) \otimes \left(\vec P(\theta_2)\times d \vec P(\theta_2)\right)
\end{eqnarray}
and in our case of piecewise constant curve with discontinuities $\Delta \vec P_k$, it becomes an ordered sum
\begin{eqnarray}
  \sum_{0\le n<m< N}\left(\vec P_m\times \Delta \vec P_m\right) \otimes \left(\vec P_n\times \Delta \vec P_n\right)
\end{eqnarray}

The discontinuity $\Delta \vec P(\theta)$ stays finite in the continuum limit $N \to \infty$. The continuum limit can be taken only after integrating(summing) the internal degrees of freedom of the fixed manifold of the Loop equations.
The imaginary part of our solution \eqref{SymSol} does not depend on the point on a circle. Therefore, it contributes a constant term into $\vec S_{m,n}$ which cancels in the difference $\vec S_{n,m}- \vec S_{m,n}$ in the exponential, as it should.

Let us look at the correlation function \eqref{CorrFunc}.
First, we expand and simplify the dot product involved
\begin{eqnarray}\label{OmegaDot}
    \vec \omega_m \cdot \vec \omega_n = \frac{-\sigma_m \sigma_n}{4} \cot ^2\left(\frac{\beta }{2}\right) 
\end{eqnarray}

The terms $S_{m,n}, S_{n,m}$ in \eqref{CorrFunc} have the following form 
\begin{eqnarray}
&& \vec \rho \cdot \vec S_{n, m} =\exp{ \imath \beta \sigma_n} A_{n, m};\\
&& \vec \rho \cdot \vec S_{m, n} =\exp{ \imath \beta \sigma_m} A_{m, n + N};\\
&& A_{n, m} = \Re R^\star \frac{\displaystyle\sum_{k=n}^{m-1} \exp{\imath \alpha_{k,n} }}{ 2 \sin(\beta/2) (m  -n)};\\
&& \alpha_{k,n} = \beta\sum_{\substack{l=0\\ l \neq n}}^{k} \sigma_l;\\
&& R = \rho_x + \imath \rho_y
\end{eqnarray}
\subsection{Critical phenomena in statistical limit}

Now we are prepared to average over spin variables $\sigma_l, l = 0\dots N-1$.
This expression singles out the variables $\sigma_n, \sigma_m$ so we can sum over these two variables, leaving the rest of $\sigma_l$ free, except for a constraint $\sum_\sigma = r q$. This constraint can be implemented as a discrete Fourier integral:
\begin{eqnarray}
    \delta[q r- \sum\sigma] = \oint \frac{ d \omega}{2 \pi} e^{  \imath \omega  \left(q r -  \sum \sigma\right)}
\end{eqnarray}

We start with 
\begin{eqnarray}
&& \frac{-1}{4} \cot ^2\left(\frac{\beta }{2}\right)\nonumber\\
&&\VEV{\sigma_m \sigma_n \oint \frac{ d \omega}{2 \pi} e^{  \imath \omega  \left(q r -  \sum \sigma\right)}\exp{\imath  \exp{ \imath \beta \sigma_n} A_{n, m} -\imath  \exp{ \imath \beta \sigma_m} A_{m, n+N} }}_{\sigma_l = \pm 1};
\end{eqnarray}

Then we take the $\omega$ integral out of the sum:
\begin{eqnarray}
&& \oint \frac{ d \omega}{2 \pi} e^{  \imath \omega q r}\nonumber\\
    &&\VEV{\sigma_m \sigma_n  e^{ -\imath \omega  \sum \sigma}\exp{\imath  \exp{ \imath \beta \sigma_n} A_{n, m} -\imath  \exp{ \imath \beta \sigma_m} A_{m, n+N} }}_{\sigma_l = \pm 1} ;
\end{eqnarray}

This expression can be readily averaged over two variables $\sigma_m, \sigma_n $, which reduces to the sum of four terms with $\sigma_n, \sigma_m = \pm 1$.
Keeping only the factors depending on $\sigma_n, \sigma_m$
\begin{eqnarray}
    &&\VEV{\sigma_m \sigma_n   e^{ -\imath \omega  (\sigma_n + \sigma_m)}\exp{\imath  \exp{ \imath \beta \sigma_n} A_{n, m} -\imath  \exp{ \imath \beta \sigma_m} A_{m, n+N} }}_{\sigma_n, \sigma_m= \pm 1} =\nonumber\\
    &&\Phi\left(A_{n, m},\omega,\beta\right) \Phi\left(-A_{m, n+ N},\omega,\beta\right);\\
    && \Phi(A,\omega,\beta) = \sin (\omega -i \sin (A \beta )) (\sin (\cos (A \beta ))-i \cos (\cos (A \beta )))
\end{eqnarray}

The next step would be averaging over the remaining variables $\sigma_l, l \neq n, l \neq m$. These variables are split into two sets: one is used in the $ A_{n, m}$, and the other is used in $ A_{m, n+N}$.
The variables $ A_{n, m}$ have a certain distribution in the statistical limit when both $\sim m \sim n \sim N \to \infty$. We also have $ \beta \sim 1/\sqrt{N} \to 0$ in that limit.

We are now considering the unconstrained distribution over $\sigma_l$, as the constraint is implemented via the discrete Fourier integral.
Let us compute the mean and variance of $ U_{n,m} =  \sum_{k=n}^{m-1} \exp{\imath  \alpha_{k,n}} $
\begin{eqnarray}
    &&\VEV{U_{n m}}  = \sum_{k=n}^{m-1} \cos^{k-1}\beta = \frac{\cos(\beta)^n - \cos(\beta)^{m}}{\cos\beta (1-\cos\beta) };\\
    && \VEV{\abs{U_{n m} -\VEV{U_{n m}}}^2 } = \sum_{l=n}^{m-1} \sum_{k=n}^{m-1} \VEV{\xi_k \xi^\star _l}=
    \sum_{l=n}^{m-1} \sum_{k=n}^{m-1} \left(\cos\beta^{|l-k|} - \cos\beta^{l}\cos\beta^{k}\right);\\
    &&  \xi_r = \exp{\imath  \alpha_{r,n}} - \VEV{\exp{\imath  \alpha_{r,n}}}
\end{eqnarray} 

In the relevant critical region $\beta \to 0, m = x /\beta^2, n = y/\beta^2, x > y$, the ratio of the variance of $U_{n,m}$ to the square of its mean tends to a finite limit
\begin{eqnarray}
    \frac{\VEV{\abs{U_{n m} -\VEV{U_{n m}}}^2 }}{\VEV{U_{n m}}^2} \to \frac{2 e^{\frac{y-x}{2}}+x-y-2}{\left(e^{-\frac{x}{2}}-e^{-\frac{y}{2}}\right)^2}-1
\end{eqnarray}
\pct{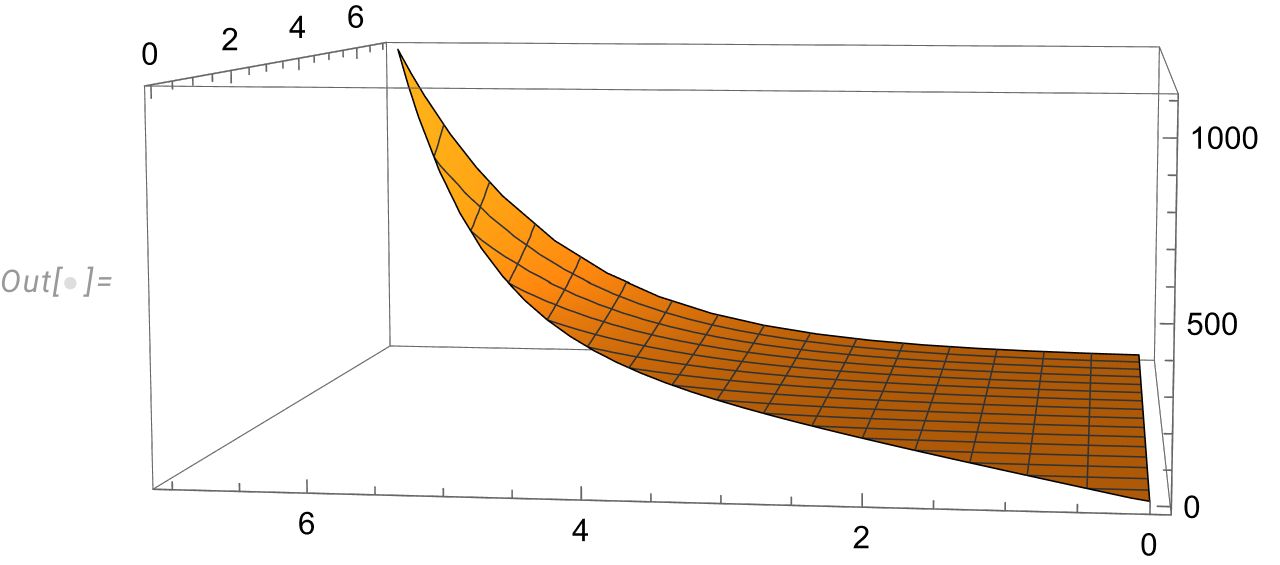}{The relative variance computed in the text in the statistical limit.}

This function looks singular, but it is positive and finite in the allowed region $0 < y < x$ ( see Fig.\ref{fig::VariancePlot}).
Therefore, the CLT does not apply to the distribution of $ A_{n, m}$, and all the moments of this distribution must be computed to obtain the probability distribution as a Mellin transform.

Calculating these moments represents \textbf{another challenge to the number theorists}. 
This complication is good news for our theory: we have critical phenomena with a non-Gaussian distribution in the statistical limit.
Still, in the next section, we derive an exact formula for the mean value of the enstrophy as a function of the chemical potential $\mu$, relating it to the functions of the number theory.

\subsection{Analytic solution for the enstrophy}

Let us find analytic formulas for observables of this remarkable statistical distribution, which is isomorphic to the Ising model tied to the ensemble of fractions.
The one-dimensional Ising models are usually solvable, and this one is no exception.
The simplest quantity is the enstrophy related to our random variables in the previous section.
Setting $ \vec r =0$ in \eqref{CorrFunc} (which is possible at finite $N$) we find the following relation
\begin{eqnarray}\label{enstrophy}
   && \VEV{\vec\omega(\vec 0)^2} = \frac{1}{ 4(t+ t_0)^2}  \sum_{0\le n<m< N}\VEV{\vec \omega_m \cdot \vec \omega_n }_{\mathbb{E}(N)}
\end{eqnarray}
where the big Euler ensemble average $\VEV{\dots}_{\mathbb{E}(N)}$ denotes averaging over $p,q,r $ and the Ising variables $\sigma$ subject to the global constraint $\sum\sigma_k = q r$ .
The general theory in section \ref{EulerEns} expressed the big Euler ensemble average in terms of the small Euler ensemble average of the average over spins.

So, we start the computation of the enstrophy by averaging over spins $\sigma_k$, which is rather simple.
Let us use the explicit expression \eqref{OmegaDot} for the dot product $\vec \omega_m \cdot \vec \omega_n $.
Now, using the one-dimensional Fourier integral for the global constraint on $\sigma$, we get the unconstrained average:
\begin{eqnarray}
    &&\VEV{\vec \omega_m \cdot \vec \omega_n }_{\sigma} =   -\frac{1}{4} \cot ^2\left(\frac{\beta }{2}\right)
    \int_{-\pi}^\pi \frac{d \omega}{2 \pi} e^{\imath \omega q r} \VEV{\sigma_m\sigma_n\exp{-\imath \omega \sum_0^{N-1} \sigma_l }}_\sigma 
\end{eqnarray}

Averaging it over $\sigma_m, \sigma_n$ we find
\begin{eqnarray}
&& \VEV{\exp{-\imath \omega (\sigma_n + \sigma_m) }\sigma_m \sigma_n}_{\sigma_m, \sigma_n} =-\sin^2(\omega)
\end{eqnarray}

Now, averaging over the remaining $\sigma_r, r \neq m,r \neq n$ is straightforward. 
\begin{eqnarray}
 \VEV{\exp{-\imath \omega \sum_{\substack{l=0 \\l \neq n, l \neq m}}^{N-1}\sigma_l }}_{\sigma} = \cos^{N-2} \omega
\end{eqnarray}

We arrive at the integral
\begin{eqnarray}
    &&\VEV{\vec \omega_m \cdot \vec \omega_n }_{\sigma} =   \frac{\cot ^2\left(\frac{\beta }{2}\right)}{4} \int_{-\pi}^\pi \frac{d \omega}{2 \pi} e^{\imath \omega q r} \cos^{N-2} \omega \sin^2 \omega
\end{eqnarray}

This integral  is calculable (see \cite{DecayTurb23}):
\begin{eqnarray}
    &&\VEV{\vec \omega_m \cdot \vec \omega_n }_{\sigma} = \frac{ 2(N-q^2 r^2)\cot ^2\left(\frac{\beta }{2}\right)}{2^{N}(N^2-q^2 r^2)} \Binom{N-2}{\frac{1}{2} (N+q r-2)}
\end{eqnarray}

Summing over $0 \leq n < m < N$ yields $N(N-1)/2$, leading to our final answer
\begin{eqnarray}\label{exact enstrophy}
    && \VEV{\vec\omega(\vec 0)^2} = \frac{1}{ (t+ t_0)^2}\VEV{2^{-N-4} S(q)  \left(N-q^2 r^2\right) \Binom{N}{(N+q r)/2}}_{\mathcal{E}(\mu)};\\
    &&  S(q) = \sum_{\substack{p=1 \\ (p,q)=1}}^{q-1}\cot ^2\left(\frac{\pi p}{q}\right);
\end{eqnarray}

We investigated this new function $S(q)$ in \cite{MigBul23} and represented it in terms of so-called multitotients \cite{multitotients}. For the reader's convenience, we present the computations leading to this representation in Appendix\ref{Totients}.
\begin{eqnarray}  \label{new sum}  
    && S(q) =\frac{\varphi_2(q)}{3} - \varphi_1(q) ;\\
    && \varphi_l(q) = q^l \prod_{p|q}\left(1 - \frac{1}{p^l}\right) ;\\
    && \varphi_1(q) = \varphi(q)
\end{eqnarray}

Here $p|q$ are prime factors of $q$. 
This remarkable identity can be directly verified using \Mathematica \cite{DecayTurb23}. It takes over a minute of CPU to compute and simplify $S(100) = 2360$. The same result using the multitotient formula takes $140$ microseconds.
Here is the table of $S(q)$ for $2 \le q \le 10$
\begin{eqnarray}
\left(
\begin{array}{cc}
 2 & 0 \\
 3 & \frac{2}{3} \\
 4 & 2 \\
 5 & 4 \\
 6 & 6 \\
 7 & 10 \\
 8 & 12 \\
 9 & 18 \\
 10 & 20 \\
\end{array}
\right)
\end{eqnarray}

This $S(q)$ is an even integer for $q >3$, and here is a simple proof.
\begin{proof}
The second term $-\varphi(q)$ in $S(q)$ is an even integer, and the first term  can be rewritten as
\begin{eqnarray}
    1/3 q^2 \prod_{p|q}\left(1 - \frac{1}{p^2}\right)  = 1/3\prod_{p|q}p^{2(\alpha_p-1)} (p^2-1);
\end{eqnarray}
where $\alpha_p\ge 1$ is a multiplicity of the prime factor $p$.  The remaining factors $(p^2-1)= (p-1)(p+1)$. The sequence of three integers $(p-1),p,(p+1)$ has a factor divisible by three, and it cannot be $p$ as it is a prime number, with the only exception being $q=p=3$. Starting with $q =4$ there is always a factor of $3$ either in $(2^2-1)$ or in $3^{2(\alpha_3-1)} $ or in any other prime factors $(p^2-1)$ analyzed above.
The  divisibility by $2$ is similar: both factors $p\pm 1$ for any prime $p>2$ are even, and any power $2^{2(\alpha_2-1)}$ with $\alpha_2 >1$ is divisible by $4$.
Therefore $p^2-1$ is divisible by $12$, so that $\frac{\varphi_2(q)}{3}$ is divisible by $4$. Thus $S(q)$ is an even integer for $q>3$.
\end{proof}

The plot of the first $100,000$ values of $S(q)$ looks as follows (see Fig.\ref{fig::SQPlot}).
\pct{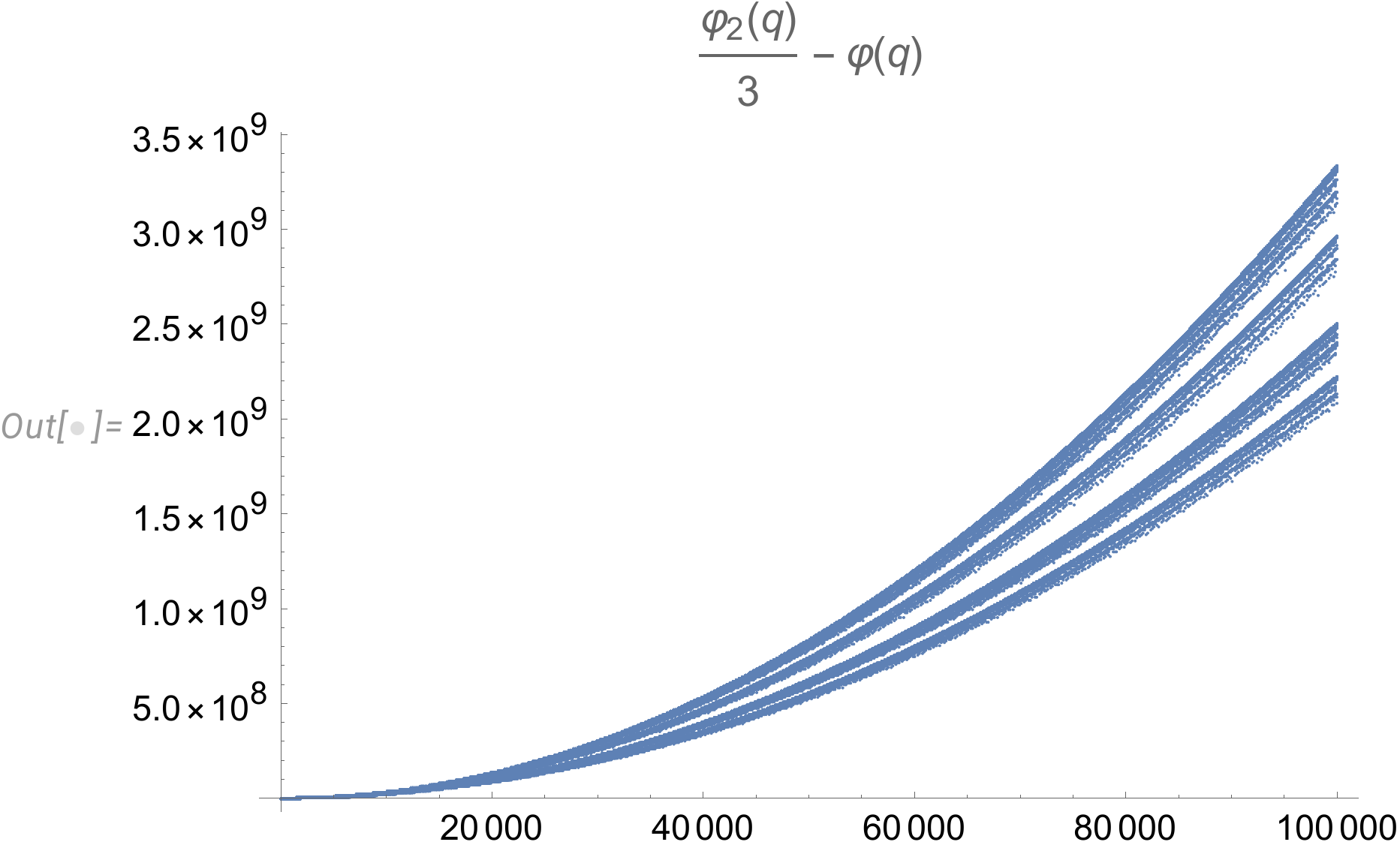}{S(q) for  $ 10 < q < 100000$. It does not reach any smooth limit; several bands persist up to infinity, similar to the Euler totient $\varphi_1(q)$.}
The appearance of prime numbers in the fluid dynamics problem is exciting; it reveals hidden relations between these different branches of modern mathematics. It reminds me of similar unexpected relations between matrix models, orthogonal polynomials, and 2D quantum gravity.

The formula \eqref{exact enstrophy} is our exact solution for enstrophy, expressing it as a calculable average over the small Euler ensemble. In the next section, we compute it in the local limit $\mu \to 0$.

\subsection{The local limit of the energy dissipation}

In the limit $ N \to \infty$, the term with zero winding $r=0$ dominates the sum $\sum_r$, which is only possible for even $N$.
Its asymptotic limit is
\begin{eqnarray}
   2^{-N-5} N\Binom{N}{N/2}\to \frac{\sqrt{N}}{16 \sqrt{2 \pi}}
\end{eqnarray}

The remaining terms, with $ r \neq 0$ can be replaced by an integral of the asymptotic expansion of the exact expression
\begin{eqnarray}
   \sqrt{N}  \int_{-\8}^{\8} d z\frac{e^{-\frac{z^2}{2}} \left(1-z^2\right)}{16 \sqrt{2 \pi }} =0
\end{eqnarray}

The leading term cancels after integration, so the $r = 0$ contribution dominates the sum.
The next  term  of expansion in $1/N$ cancels as well
\begin{eqnarray}
   \int_{-\8}^{\8} d z \frac{e^{-\frac{z^2}{2}} \left(z^2-1\right) \left(z^4-6 z^2+3\right)}{192 \sqrt{2 \pi } \sqrt{N}} =0
\end{eqnarray}

We do not need this term $O(1/\sqrt{N})$, as the dominant $r=0$ term grows as $\sqrt{N}$. As seen in the section \ref{EulerEns}, the remaining terms decrease faster than exponential with $q$, making it a nontrivial exercise in number theory to find an analytic formula for the partition function and the enstrophy in the odd Euler ensemble.

Let us concentrate on even $N$, as this term dominates the grand canonical ensemble we study.
\begin{eqnarray}
   && (t+ t_0)^2\VEV{\vec\omega(\vec 0)^2} \to \frac{1}{2}\frac{\displaystyle \int_0^\infty d N \sqrt{N} e^{-\mu N}\displaystyle\sum_{2<q<N} S(q)}{16 \sqrt{2\pi }Z(\mu)} 
\end{eqnarray}

The extra factor of $1/2$ came from skipping all the odd values of $N$. The remaining sum over even $N$ tends to be half the asymptotic expression's integral for these even $N$.
The asymptotic behavior of multitotient summators has been known for a century \cite{multitotients}
\begin{eqnarray}
    \sum_{m=2}^N\varphi_l(m)  \to \frac{ N^{l+1}}{ (l +1) \zeta(l+1)}
\end{eqnarray}
where $\zeta(n)$ is the Riemann's zeta function.
We obtain the following local limit of the energy decay in the grand canonical Euler ensemble
\begin{eqnarray}\label{finalAnswer}
    && \partial_t E = - \nu \VEV{\vec\omega(\vec 0)^2}  = -\frac{\mathcal B \nu}{\mu^2(t+ t_0)^2} ;\\
    \label{AnomalousConstant}
    && \mathcal{B} = \frac{35 \pi^2 }{3456  \zeta (3)} = 0.08315129725;
\end{eqnarray}

This energy dissipation stays finite in a turbulent limit $\nu \to 0$ provided
\begin{equation}
    \mu \propto \sqrt{\nu} \to 0
\end{equation}

Let us now estimate the odd $N$ contribution to the enstrophy. Here, the sum of the spin average is dominated by $q \ll N$, where we cannot use the asymptotic formula of the number theory for totients.
Thus, we used numerical results of direct computations of the small Euler ensemble contribution from odd $N$ to the enstrophy 
\begin{eqnarray}
  &&\displaystyle\sum_{\text{odd }N}e^{-\mu N}\displaystyle\sum_{ \substack{2 <  q < N\\ \text{odd } q}}\left(\frac{\varphi_2(q)}{3}-\varphi(q)\right)\displaystyle\sum_{\substack{r>0\\\text{odd } r}}2^{-N-4}  \left(N-q^2 r^2\right) \Binom{N}{(N+q r)/2}
\end{eqnarray}

This contribution $\Delta E$ comes out negative.
We fitted these results as $\log(- \Delta E )\approx a + b \log N + c \log \log N $ (see Fig.\ref{fig::FitEnsOddN}).
\pct{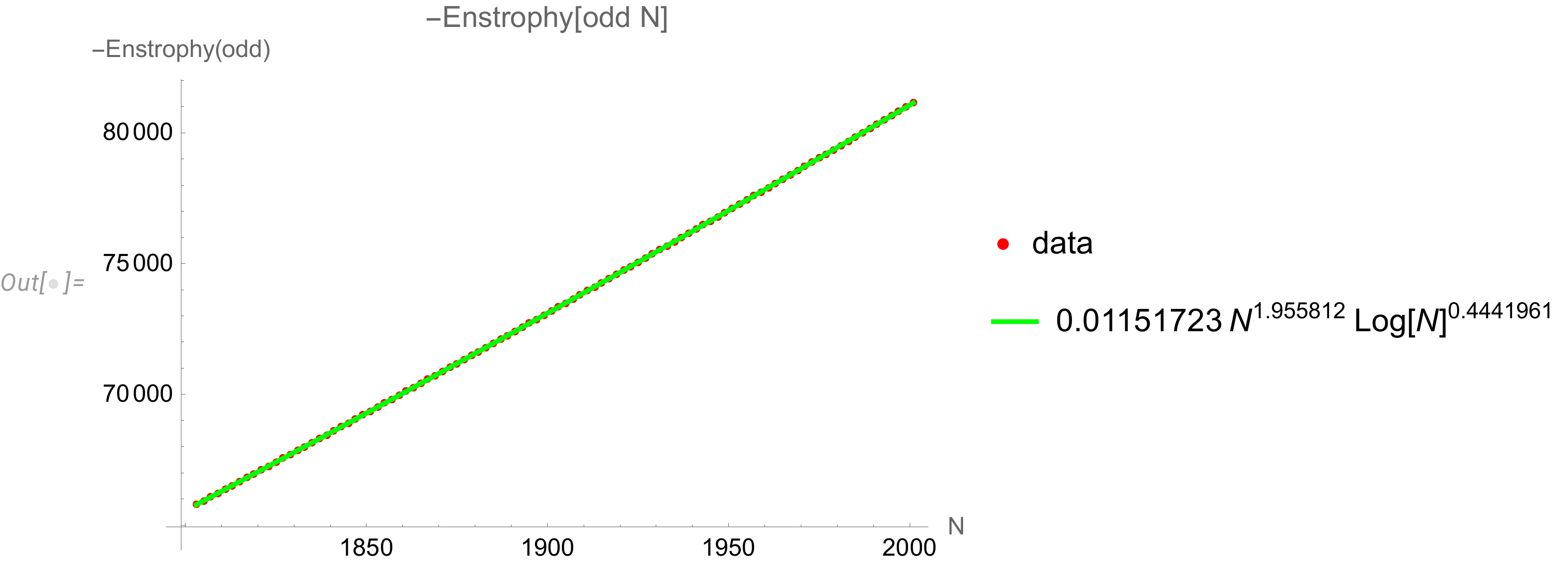}{The direct computation of the odd $N$ contribution $\Delta E$ to the enstrophy with $20$ digits  fitted as $\Delta E \approx -  e^a N^b \log ^c N$.}
Here are the fit statistics
\begin{eqnarray}
\begin{array}{|lllll}
\hline
 \text{} & \text{Estimate} & \text{Standard Error} & \text{t-Statistic} & \text{P-Value} \\
 1.0000 & -4.4639 & 0.00012141 & -36768. & 2.6011*10^{-348} \\
 x & 1.9558 & 0.000015743 & 124230. & 1.333*10^{-399} \\
 \log (x) & 0.44420 & 0.00011885 & 3737.3 & 5.3357*10^{-252} \\
\end{array}
\end{eqnarray}

We have the following estimate for  the odd $N$ contribution to the enstrophy in the grand canonical Euler ensemble in the local limit:
\begin{eqnarray}
    \frac{\Delta E}{Z(\mu)} \approx -\frac{0.068618 \log ^{0.44420}\left(\frac{1}{\mu }\right)}{\mu ^{0.45581}}
\end{eqnarray}

The leading term grows  $1/\mu^2$; therefore, this correction is negligible.

The enstrophy divergence corresponds to anomalous dissipation in our theory.
This divergence is the dual version of the original anomalous dissipation  in the \NS{}
velocity field, coming from singular vorticity regions \cite{M23PR, M22}(Burgers vortexes).
Here, it comes from large fluctuations of the fractal curve in the grand canonical Euler ensemble. These are quantum effects related to the prime factorization of large integers.

\subsection{The higher moments of the enstrophy}

The higher moments of the distribution of enstrophy are also calculable \cite{DecayTurb23}. 
We take the $2 n$ point correlation function in the limit of the vanishing loop
\begin{eqnarray}\label{enstrophyMoments}
   && \VEV{\omega(\vec 0)^{2n}} = \frac{1}{ 4^n(t+ t_0)^{2 n}}  \sum_{0\le m_1<\dots m_{2n} < N}\VEV{\omega_{m_1} \dots  \omega_{m_{2n}} }_{\mathbb{E}(N)};\\
   && \omega_{k} = \frac{\imath \sigma_k}{2} \cot\left(\frac{\beta}{2}\right)
\end{eqnarray}

With all different $m_1<\dots <m_{2 n} $, the averaging over the Ising variables $\sigma_l$ is straightforward. It leads to the integral over $\omega$
\begin{eqnarray}
    J(n,q r, N) = \oint \frac{d \omega}{2 \pi} \exp{ -\imath q r} \cos^{N - 2 n} (\omega) \sin^{2 n} (\omega) 
\end{eqnarray}

This integral is reduced to the hypergeometric function in \cite{DecayTurb23}:
\begin{eqnarray}
    &&J(n,q r, N) =2 (-1)^n \Binom{N-2 n}{\frac{1}{2} (N-q r)} \nonumber\\
    && \, _2F_1\left(-2 n+N+1,\frac{1}{2} (N+q r+2);\frac{1}{2} (-4 n+N+q r+2);-1\right)
\end{eqnarray}

In particular,
\begin{eqnarray}
    &&J(n,0, N) =\frac{(-1)^n \Gamma \left(\frac{1}{2} (-2 n+N+1)\right)}{\Gamma \left(\frac{1}{2}-n\right) \Gamma \left(\frac{N}{2}+1\right)}
\end{eqnarray}

This term does not depend on $q$; it dominates in the local limit $N \to \infty$ at fixed $n$.

The general formula for the numerator of the moments at fixed $N$ reads
\begin{eqnarray}
   && \VEV{\omega(\vec 0)^{2n}}_{\mathcal{E}(N)} \propto \Binom{N}{2 n} (4 t)^{-2 n} \sum_{2 <q< N} S(n,q) \sum_{\substack{r = -\floor{N/q}\\ 2|(N-q r)}}^{\floor{N/q}} J(n, q r, N);\\
   &&S(n,q) = \sum_{\substack{p=1\\(p,q)=1}}^{q-1}\cot^{2n}\left(\frac{\pi p}{q}\right);
\end{eqnarray}

This cotangent sum is reduced to the superposition of multi-totients in Appendix.\ref{Totients}
\begin{eqnarray}
    &&S(n,q) = (-1)^n \varphi(q) - (-4)^n  \sum _{j=1}^n \frac{B_{2 j} \varphi_{2 j}(q)\text{BernSum}(n,n-j)}{(2 j)!};\\
    && \varphi_l(m) = m^l \prod_{p|m}\left(1 - \frac{1}{p^l}\right) ;
\end{eqnarray}

The rational coefficients $\text{BernSum}(n,n-j)$  are given by recurrent relations 
\begin{eqnarray}
&&\text{BernSum}(0,0) =1;\\
&&\text{BernSum}(n,m) = 0 \text{ if } n < m;\\
&& \text{BernSum}(n,m)=\sum _{j_1=0}^m \sum _{j_2=0}^{m-j_1} \frac{B_{2 j_1} B_{2 j_2} \text{BernSum}(n-1,m-j_1-j_2)}{(2 j_1)! (2 j_2)!};
\end{eqnarray}

The specific cases are considered in Appendix\ref{Totients}.
We are interested in the local limit, even $N \to \infty$.
In this limit, the highest totient $\varphi_{2n}(q)$ dominates the sum, and we find for the sum 
\begin{eqnarray}
 && \VEV{\omega(\vec 0)^{2n}}_{\mathcal{E}(\mu)} =   \frac{\displaystyle\sum_{ even N} \exp{- \mu N}\Omega(N,n)}{(t+ t_0)^{2 n}Z(\mu)};\\
&&\Omega(N,n) = -\frac{\left(-\frac{1}{4}\right)^n B_{2 n} N^{2 n+1} \Gamma \left(n+\frac{1}{2}\right) \Binom{N}{2 n} \Gamma \left(-n+\frac{N}{2}+\frac{1}{2}\right)}{\pi  \zeta (2 n+1) \Gamma (2 n+2) \Gamma \left(\frac{N}{2}+1\right)}
\end{eqnarray}

Finally, in the limit $\mu \to +0$, replacing the sum over even $N$ by $1/2$ of the integral of the asymptotic of this product of gamma functions, we find
\begin{eqnarray}
    &&\VEV{\omega(\vec 0)^{2n}} \to  \frac{\Xi_n}{(t+t_0)^{2 n}\mu ^{3 n-1}}\; \text{ if } n >0;\\
    && \Xi_n = \frac{\pi\sqrt{\pi } (-1)^{n-1} 2^{2-3 n} B_{2 n}  \Gamma \left(3 n+\frac{3}{2}\right) }{9 \zeta (2 n+1) \Gamma (n+1) \Gamma (2 n+2) }
\end{eqnarray}

These coefficients $\Xi_n$ are all positive. Here are the first five values
\begin{eqnarray}
\begin{array}{cc}
\hline
 \text{Exact} & \text{Numerical} \\
   \frac{35 \pi ^2}{3456 \zeta (3)}, &0.08315129725 \\
   \frac{1001 \pi ^2}{983040 \zeta (5)},&0.009692016257 \\
   \frac{230945 \pi ^2}{528482304 \zeta (7)}, &0.004277271996\\
  \frac{185910725 \pi ^2}{463856467968 \zeta (9)},&0.003947746107 \\
   \frac{15193976525 \pi ^2}{24189255811072 \zeta (11)},&0.006196324003 \\
\end{array}
\end{eqnarray}

\subsection{The vorticity distribution}

Using the analytic formulas for the moments of the enstrophy, we can study nonperturbative effects in our solution, such as the PDf of the vorticity\cite{DecayTurb23}.
Let us consider the Fourier transform of the PDF corresponding to the vorticity moments, 
\begin{eqnarray}
    H( \vec J, \mu, t) = \VEV{ \exp{ \imath \vec J \cdot \hat \Omega\cdot \vec \omega(0)}}_{\hat \Omega\in O(3),\mathbb E(\mu)} 
\end{eqnarray}

The averaging over the global rotation matrix $\hat \Omega\in O(3)$ yields
\begin{eqnarray}
     H( \vec J, \mu, t) =\VEV{ \frac{\sin \left(\sqrt{\vec J^2 \vec \omega(0)^2}\right)}{\sqrt{\vec J^2 \vec \omega(0)^2}}}_{\mathbb E(\mu)}  = \sum_{n=0}^\infty \frac{(-\vec J^2)^n}{(2 n +1)!}\VEV{\vec \omega(0)^{2 n} }_{\mathbb E(\mu)}
\end{eqnarray}

Using our solution for the moments, we get a universal scaling function
\begin{eqnarray}
  && H( \vec J, \mu, t)  = 1 + \mu \mathcal F\left(\frac{\vec J^2}{(t + t_0)^2 \mu^3}\right);\\
  && \mathcal F(z) =  \sum_{n=1}^\infty \frac{z^n (-1)^n\Xi_n }{(2 n +1)!}
\end{eqnarray}

This expansion for the scaling function has a finite radius of convergence, as it follows from the asymptotic
\begin{eqnarray}
   &&\frac{\Xi_n }{(2 n +1)!}  \to \frac{\pi } {3 n^2 R^n};\\
   &&R = \frac{128 \pi ^2}{27};
\end{eqnarray}

The singularity is located at negative $z = -R$.
The expansion can be truncated inside the convergence region $|z| < R$. Here is the corresponding plot for expansion truncated at $n= 200$ (see Fig.\ref{fig::Fz200}).
\pct{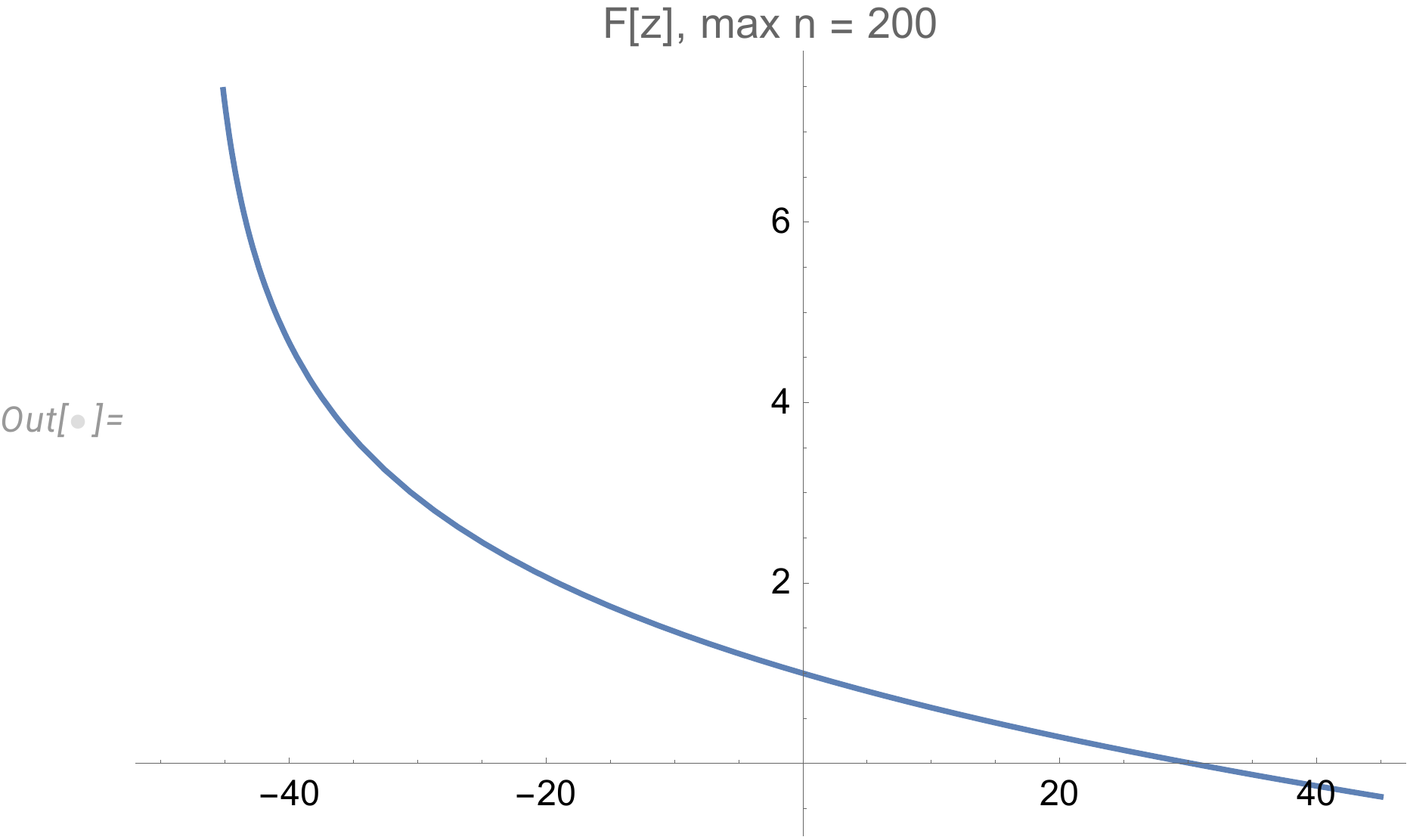}{The scaling function $F(z), z =\frac{\vec J^2}{(t + t_0)^2 \mu^3}$, truncated at $n = 200$.}
At large positive $z$, it behaves as 
\begin{eqnarray}
    &&\mathcal F(z) \sim \pi/3\sum _{n=1}^{\infty } \frac{(-z/R)^n}{n^2} = \pi/3\text{Li}_2\left(-\frac{1}{27} \left(128 \pi ^2 z\right)\right)  \sim -\frac{\pi}{6} \log ^2(z)
\end{eqnarray}

This singularity of the Fourier transform at the imaginary axis corresponds to the  exponential decrease of the PDF
\begin{eqnarray}
    &&W(|\vec \omega(0)| > x) \propto \exp{- a x (t + t_0) \mu^{\sfrac{3}{2}}};\\
    && a = \sqrt{R} = \frac{8}{3} \sqrt{\frac{2}{3}} \pi ;
\end{eqnarray}

There is no continuum limit for the distribution of vorticity. 
However, this statistical system is renormalizable in the sense that the redefinition of the source 
\begin{eqnarray}
    \vec j_R = \mu^{-\sfrac{3}{2}} \vec j
\end{eqnarray}
eliminates the singularity in the same way as the redefinition of viscosity
\begin{eqnarray}
    \nu_R = \mu^{-2} \nu
\end{eqnarray}
eliminated the singularity in the energy dissipation.

\section{The decay index spectrum}

The vicinity of the fixed point in nonlinear dynamic systems provides the most interesting physical parameters, such as anomalous dimensions of various local operators in the theory of renormalization group.
Our system is no exception. 

\subsection{Linearized loop equation}
Let us perturb the momentum loop equation\eqref{PloopEq} and study the general properties of this perturbation $\delta \vec P(t,\theta) = \vec Q(t,\theta)$.
We get a linearized equation for $\vec Q(t,\theta)$ of the form
\begin{eqnarray}
   \nu \partial_t \vec Q = \hat H_1[\vec P] \cdot \vec Q + \hat H_2[\vec P] \cdot \Delta \vec Q
\end{eqnarray}

These matrices $\hat H_{1,2}[\vec P]$ for the decaying solution \eqref{decayingSolution} have an explicit time dependence 
\begin{eqnarray}
    \hat H_{1,2}[\vec P]  = \frac{\nu}{2(t + t_0)\gamma^2} \hat H_{1,2}[\vec F]
\end{eqnarray}
which follows from the fact that the RHS of the original loop equation \eqref{PloopEq} represents the third-degree homogeneous functional of $\vec P$.
We  find the linear equation of the form 
\begin{eqnarray}
    (t + t_0)\partial_t \vec Q = \hat L_1  \cdot \vec Q + \hat L_2 \cdot \Delta \vec Q
\end{eqnarray}
with $L_{1,2} = \oh \hat H_{1,2}[\vec F]$ being some tensors in $\mathbb{R}_d$. There is no explicit $\theta$ dependence other than through the fixed point solution $\vec F(\theta)$. We shall write explicit equations in a minute.
This equation has power-like solutions
\begin{eqnarray}
    \vec Q(t,\theta) = (t+ t_0)^{-\lambda} \vec G(\theta)
\end{eqnarray}
with the index $\lambda$ determined from the eigenvalue problem
\begin{eqnarray}
   &&- \lambda \vec G = \hat L_1  \cdot \vec G + \hat L_2 \cdot \Delta \vec G
\end{eqnarray}

We found scaling laws with anomalous dimensions without any renormalization group. We studied linearized equations near the fixed point. In the vicinity of a conventional fixed point, the perturbations would decay or grow exponentially with time by the Lyapunov indexes of the linearized equation.
In our decay fixed point, the base solution decays as $1/\sqrt{t}$, making the linearized operator decay $\propto 1/t$. This decay of the operator converts the exponential decay of perturbations into a power decay without having scale invariant field theory.

We can go one step forward before specifying the parameters of this equation.
Let us use our polygonal approximation for $\vec F, \vec G$. This equation becomes a recurrent equation
\begin{eqnarray}
&& \vec G \Rightarrow \frac{\vec G_{k+1} + \vec G_{k}}{2};\\
&& \Delta \vec G \Rightarrow \vec G_{k+1} - \vec G_{k};\\
\label{ABGeq}
    &&(\lambda \hat I +\hat A_k) \cdot \vec G_{k+1} = (-\lambda \hat I +\hat B_k )\cdot \vec G_k;
\end{eqnarray}

We can solve it as a matrix product (in reverse order)
\begin{eqnarray}\label{GkSol}
   && \vec G_{k+1} =  \hat M_k(\lambda) \cdot \vec G_0;\\
   && \hat M_k(\lambda) = \prod_{i=k}^{i=0} (\hat I \lambda +\hat A_k)^{-1}(-\hat I\lambda +\hat B_k )
\end{eqnarray}

The periodicity requires $G_N = G_0$ which leads to the eigenvalue equation (this is already $d\otimes d$ matrix equation)
\begin{eqnarray}
    &&\hat M_N(\lambda) \cdot \vec G_0 = \vec G_0
\end{eqnarray}

As a result, we arrive at the spectral equation for the fractal dimensions of decaying turbulence
\begin{eqnarray}\label{spectralEq}
    \det \left[\hat M_N(\lambda) - \hat I\right] =0
\end{eqnarray}

Here is the explicit form of these two matrices  ( with $\Delta \vec F_k = \vec F_{k+1} - \vec F_k$,  see \cite{DecayTurb23})
\begin{eqnarray}
    &&2\hat A_k  = \gamma \lambda \hat I -(\gamma -2 \imath) \Delta \vec F_k\otimes \Delta \vec F_k +(2 \gamma +3 \imath) \Delta \vec F_k\otimes\vec F_k+\imath \vec F_k \otimes\Delta \vec F_k;\\
    && 2\hat B_k = - \gamma \lambda\hat I + \gamma  \Delta \vec F_k\otimes \Delta \vec F_k+(2 \gamma +\imath) \Delta \vec F_k\otimes\vec F_k -\imath \vec F_k \otimes\Delta \vec F_k
\end{eqnarray}

Note that these two matrices are functions of $\gamma$, unlike the fixed point solution \eqref{SymSol},\eqref{alphaEq}, where the $\gamma$ dependence dropped.
This fact will be important for the distribution of the velocity circulation.

The spectral equations are simpler than they look.
The key to simplification is a simple algebra satisfied by the two vectors $\Delta \vec F_k, \vec F_k$ for arbitrary $k$
\begin{eqnarray}
    &&\Delta \vec F_k \cdot \vec F_k = -\oh;\\
    && \Delta \vec F_k \cdot \Delta \vec F_k = 1;\\
     && \vec F_k \cdot \vec F_k = \oq;
\end{eqnarray}

Using this algebra, we may expand the inverse matrix $(\hat I \lambda +\hat A_k)^{-1}(-\hat I \lambda+\hat B_k)$ on five various tensor products and solve linear equations for expansion coefficients. We get (see \cite{DecayTurb23})
\begin{eqnarray}
    &&(\hat I \lambda +\hat A_k)^{-1}(-\hat I \lambda+\hat B_k) = \nonumber\\
    &&\frac{\mu_0 \hat I+ \mu_1\vec F_k \otimes\Delta \vec F_k + \mu_2 \Delta \vec F_k \otimes\vec F_k + \mu_3\Delta \vec F_k \otimes\Delta \vec F_k + \mu_4\vec F_k \otimes\vec F_k}{4\lambda^2 -\gamma^2 };\\
    && \vec \mu = \left\{\gamma ^2(1-4 \lambda ^2),2,2 (\gamma +i) (2 \gamma(1 +2 \lambda) -i),2 i \gamma(1 +2 \lambda) +1,4-4 i \gamma \right\}
\end{eqnarray}

Thus, we have factored out the poles from the matrix $\hat M_k$
\begin{eqnarray}
    && \hat M_n =\frac{\hat H_n}{((4 \lambda^2-1)\gamma^2)^{n} };\\
    && \hat H_n = \prod_{k=n}^{k=0} \left(\mu_0 \hat I+ \mu_1\vec F_k \otimes\Delta \vec F_k + \mu_2 \Delta \vec F_k \otimes\vec F_k + \mu_3\Delta \vec F_k \otimes\Delta \vec F_k + \mu_4\vec F_k \otimes\vec F_k\right);\\
    && \vec G_k = ((4 \lambda^2-1)\gamma^2)^{N-k} \hat H_k \vec g_0;\\
    &&\hat H_N \vec g_0 = ((4 \lambda^2-1)\gamma^2)^{N} \vec g_0;
\end{eqnarray}
$\hat M_n$ has a $n$-th degree poles at $\lambda = \pm \frac{\gamma}{2}$ and no other poles. The spectrum is determined by the polynomial equation of degree $6 N$
\begin{eqnarray}\label{spectralEqPolynomial}
    \text{spectrum}: \det \left[\hat H_N - \hat I ((4 \lambda^2-1)\gamma^2)^{N}\right] =0
\end{eqnarray}

In three dimensions, this equation can be written as a cubic characteristic polynomial equation
\begin{eqnarray}
    &&\det \hat H_N + x \frac{\tr \hat H_N^2 - (\tr \hat H_N)^2}{2} + x^2 \tr \hat H_N - x^3 =0;\\
    && x= ((4 \lambda^2-1)\gamma^2)^{N}
\end{eqnarray}
\subsection{The spectral identity and Wilson loop asymptotics}
The Wilson loop \eqref{WilsonLoop} doubles as a Fourier transform of the PDF for the velocity circulation. This PDF can be obtained by inverse Fourier transform
\begin{eqnarray}
    W[C,\Gamma] = \VEV{ \delta\left(\Gamma - \oint_C d \vec r \cdot \vec v(\vec r)\right)} = \int_{-\infty}^\infty \frac{d \gamma}{2 \pi \nu} \exp{-\imath \frac{\gamma}{\nu} \Gamma}\Psi[\gamma ,C] ;
\end{eqnarray}

However, substituting the leading term of the solution of the loop equation into this general formula leads to a paradox: this leading term does not depend on $\gamma$. Formally, we get the delta function $\delta(\Gamma)$, which means that this leading term is not sufficient to get the PDF; we need the next correction.

We have found the linear correction to the fixed point, and now we can fix these formulas. Expanding the correction and keeping the leading linear term, we find:
\begin{eqnarray}\label{CircPDF}
   && \delta W[C,\Gamma] = \frac{\imath}{\nu} \int_{-\8}^\infty \frac{d \gamma}{2 \pi} \exp{-\imath \frac{\gamma}{\nu} \Gamma}\delta\Psi[C,\gamma] ;\\
   && \delta\Psi[C,\gamma]= \VEV{\exp{ \frac{\imath \sum_k \Delta\vec C_k\cdot  \vec F_k}{\sqrt{2 \nu(t+t_0) }}}\sum_{\lambda \in \text{spectrum}}\frac{\sum_k \Delta\vec C_k \cdot \vec G_k(\lambda)}{\nu(t+t_0)^{ \lambda}}}_{\mathbb{E}(N)};
\end{eqnarray}

The distribution of the eigenvalues $\lambda$ can be studied using the resolvent
\begin{eqnarray}\label{resolvent}
    R_N(\gamma,\lambda) =  \tr{\left(\hat H'_N(\lambda)- 8 N\lambda\gamma^2((4 \lambda^2-1)\gamma^2)^{N-1}  \right)\left(\hat H_N(\lambda )-\hat I((4 \lambda^2-1)\gamma^2)^{N}\right)^{-1}}
\end{eqnarray}

By construction, at finite $N$, this is a rational function of $\lambda$ with simple poles at the spectrum,
as it follows from the representation
\begin{eqnarray}
    &&R_N(\gamma,\lambda) = \pp{\lambda}  \log \det \left(H_N(\lambda)-\hat I((4 \lambda^2-1)\gamma^2)^{N}\right) ;
\end{eqnarray}

The asymptotic behavior of this rational function at infinity (for even $N$) follows from \eqref{GkSol}:
\begin{eqnarray}
    && \hat M_N(\lambda \to \infty ) \to \hat I + \frac{\hat \Sigma}{ \lambda}  ;\\
    && \hat \Sigma = -\sum_k  (\hat A_k + \hat B_k);\\
    &&  R_N(\gamma,\lambda\to\infty) \to \nonumber\\
    &&\tr\left(- 8 N\lambda\gamma^2((4 \lambda^2-1)\gamma^2)^{-1}  \right)\left(M_N(\lambda)-\hat I\right)^{-1} \to -2 N \tr(\Sigma^{-1})
\end{eqnarray}

Combining these two properties, we get the sum over the spectrum plus an irrelevant constant
\begin{eqnarray}
    &&R_N(\gamma,\lambda) = \sum_{z \in \text{spectrum}} \frac{1}{\lambda - z} -2 N \tr(\Sigma^{-1});\\
    && \text{spectrum}: z_1,\dots  z_{6 N -3} , \det \left(\hat M_N(z_k)-\hat I\right)=0;
\end{eqnarray}

The number of zeros $z_i$ can be counted from the rational function 
\begin{eqnarray}
    \pp{\lambda} \log \det \left( \hat M_N(\lambda) - \hat I \right) = \sum_z\frac{1}{\lambda - z}  - \sum_p\frac{1}{\lambda - p}
\end{eqnarray}

At infinity, this becomes (with $ 3 N$ poles at $\lambda = -\oh \gamma$ and $ 3 N$ poles at $\lambda = +\oh \gamma$)
\begin{eqnarray}
    \frac{\#z - \#p}{\lambda} = \frac{\#z - 6 N}{\lambda}
\end{eqnarray}

On the other hand, we can write this rational function as
\begin{eqnarray}
    &&\pp{\lambda} \log \det \left( \hat M_N(\lambda) - \hat I) \right) = \tr\log \left(\hat M_N'(\lambda) (\hat M_N(\lambda) - \hat I)^{-1}\right) \nonumber\\
    &&\to \tr \left((-\Sigma /\lambda^2) /(\Sigma/\lambda) \right) = -\tr \hat I/\lambda = -\frac{3}{\lambda}
\end{eqnarray}
which yields
\begin{eqnarray}
    \#z = 6 N -3
\end{eqnarray}

Consider the anticlockwise contour $\omega$ surrounding the whole spectrum of poles $z_i$ in a complex plane.
Then, we can use the unit residues at $z_i$ to compute the contour integral, with arbitrary holomorphic function $F(\lambda)$
\begin{eqnarray}
    \int_\omega \frac{d \lambda}{2 \pi \imath} R_N(\lambda) F(\lambda) = \sum_{z \in \text{spectrum}} F(z)
\end{eqnarray}

The left side of this identity can be inserted inside the statistical average, such as the circulation PDF integral:
\begin{eqnarray}\label{WilsonLoopPDF}
   && \delta W[C,\Gamma] = \VEV{\exp{ \frac{\imath \sum_k \Delta\vec C_k\cdot  \vec F_k}{\sqrt{2 \nu(t+t_0) }}}  J(t) }_{\mathbb{E}(N)};\\
   && J(t) = \int_{-\infty}^\infty \frac{d \gamma}{2 \pi} \exp{-\imath \frac{\gamma}{\nu} \Gamma}\int_\omega \frac{d \lambda}{2 \pi } R_N(\gamma,\lambda) \frac{\sum_k \Delta\vec C_k \cdot \vec G_k(\lambda)}{\nu(t+t_0)^{ \lambda}}
\end{eqnarray} 

This integral, by design, equals the sum over the decay spectrum and will be dominated at large times by the zero $z_0$ with the smallest real part.
After averaging over the random parameters of the Euler ensemble, the spectrum will likely become continuous. However, this relation can still be used to define the large-time behavior of the Wilson loop.

In this case, it is appropriate to deform the integration contour to the steepest descent from the saddle point. This contour will allow us an analytic investigation of the large-time asymptotic.
The resolvent poles will become discontinuous across the cut in the complex $\lambda$ plane.  The presumed finite density of these condensed poles will result in logarithmic factors modifying the power laws of the time decay.
The asymptotic behavior of $W[C,\Gamma]$ at a large time $t$ is determined by the saddle point equations for two variables $\gamma, \lambda$
\begin{eqnarray}
    &&\log t =  \VEV{\pbyp{\log R_N(\gamma,\lambda)}{\lambda}}_{\mathbb{E}(N)} + \VEV{\pbyp{\log \sum_k \Delta\vec C_k \cdot \vec G_k}{\lambda}}_{\mathbb{E}(N)};\\
    && \frac{\imath \Gamma}{\nu} = \VEV{\pbyp{\log R_N(\gamma,\lambda)}{\gamma}}_{\mathbb{E}(N)} + \VEV{\pbyp{\log \sum_k \Delta\vec C_k \cdot \vec G_k}{\gamma}}_{\mathbb{E}(N)};
\end{eqnarray}

The stability of our fixed-point solution requires the condition of the saddle point $\gamma_c, \lambda_c$
\begin{eqnarray}
    \Re \lambda_c \ge 0
\end{eqnarray}

We plan to study the spectrum of fractal dimensions and asymptotic distribution of the circulation in the forthcoming paper \cite{MigBul23}, using the NYU AD supercomputer cluster.

\section{Discussion}
\subsection{Singular solutions of the \NS{} equation?}

The issue of "finite time singularity" of the \NS{} equation, particularly that without random forcing, has attracted much interest from mathematicians.
The Millennium Prize Problem of proving or disproving the smoothness in the solution of the \NS{} equation remains unsolved.
The most advanced research in this field so far has been performed by Tao (see his recent review \cite{Tao2019}). Based on a simplified model of the averaged \NS{}, Tao conjectured \cite{Tao2016} that irregular behavior occurs in finite time.
One of his arguments is based on anomalous dissipation, coming from divergent enstrophy due to singular velocity gradients. The anomalous dissipation occurs only in the vanishing viscosity limit in the \NS{} equation.

We investigated anomalous dissipation in the previous papers \cite{M23PR, M22}, where we have found the anomalous terms in the Euler Hamiltonian related to the Burgers vortex. This vortex corresponds to a singular Euler solution, with vorticity becoming the delta function at the infinitely thin vortex line in the $\nu \to 0$  limit of the \NS{} equation.
The analysis in \cite{M23PR, M22} says nothing about the finite time singularity; this solution was suggested there as a stationary solution of the \NS{} equation in the limit of vanishing viscosity without specifying the time evolution preceding this stationary solution.
In the dual theory developed in this paper, the anomalous dissipation comes from large fluctuations of our fractal curve, leading to divergent enstrophy.

Surprisingly, we have more than anomalous dissipation in the present theory: the singularity we have found exists at finite viscosity, in the spirit of Tao's conjecture. However, we are not studying a particular solution with a finite time blow-up.
On the contrary, we have a time evolution in our solution (time decay) such that vorticity distribution is singular at every moment. 
Again, let us stress it: we are not claiming any theorems about finite time singularities of solutions of \NS{} equation with some initial data. 

\subsection{Stochastic solution of the \NS{} equation and ergodic hypothesis}

Statistical "analysis of circulating or turbulent fluids"  was defined by Feynman \cite{Feynman} as the last unsolved problem of classical physics. 
We are pursuing this problem by finding a stochastic solution of the \NS{} equation covering a certain manifold. Our singularities arise in correlation functions after averaging over this manifold of solutions. 
We find this manifold (Euler ensemble) by solving the loop equation (a subset of the Hopf functional equation for the generating functional of velocity field probability). None of the particular solutions in this manifold has a finite time blow-up. The singularity emerges from averaging the distribution of these solutions over the Euler ensemble.

We take the most natural invariant measure from the point of the number theory: each element of the Euler ensemble enters with equal weight. We call it our ergodic hypothesis.
This hypothesis is not necessary to solve the loop (i.e., Hopf) equation, as any linear superposition of the found solutions would satisfy the loop equation.
The singularities of our Euler grand canonical ensemble at $\mu \to 0$  remain in local variables such as enstrophy and its PDF, indicating an intrinsic problem of the \NS{} equation. It has to be regularized at small distances, and viscosity does not provide enough regularization in our solution.

The situation reminds that of the QED in the mid-twentieth century. The continuum limit of the theory showed unexpected divergent integrals due to an infinite number of local degrees of freedom of a continuous electromagnetic field. 
In both cases, the continuum theory was an idealization of the real physical system: in the case of QED, there were other forces at small distances, eventually merging QED into the Standard Model, which is still inconsistent at Planck's ultra-small distances where the quantum gravity enters the game.
In the case of the \NS{} equation, it ignores the molecular structure of fluid. The incompressibility is also an idealization: at large gradients of velocity, the sound waves related to compressibility lead to some physical cutoff of infinite vorticity.

In other words, the \NS{} equation has limited applicability in the physical world and needs to be modified at large gradients. After all, this is a phenomenological equation resulting from the truncated expansion of friction forces in velocity gradients.
The common presumption that "viscosity regularizes the velocity field" must be tested beyond the perturbation expansion, which we did in this work.
We present a singular decay solution of the \NS{} loop equations in arbitrary dimension $d>2$. However, we did not prove that the solution of the \NS{} equation as a PDE with smooth initial data would eventually approach our stochastic solution as an asymptotic regime.

\subsection{The physical meaning of the loop equation and dimensional reduction}

The long-term evolution of Newton's dynamical system with many particles eventually covers the energy surface (microcanonical ensemble). The ergodic hypothesis (accepted in Physics but still not proven mathematically) states that this energy surface is covered uniformly.
The Turbulence theory aims to find a replacement of the microcanonical ensemble for the \NS{{} equation. This surface would also take part in the decay in the pure \NS{} equation without artificial forcing. 
In both cases, Newton and \NS{}, the probability distribution must satisfy the Hopf equation, which follows from the dynamics without specification of the mechanism of the stochastization.
Indeed, the Gibbs, as well as the microcanonical distributions in Newton's dynamics, satisfy the Hopf equation in a rather trivial way: It reduces to the conservation of the probability measure (Liouville theorem), which suggests the energy surface as the only additive integral of motion to use as a fixed point manifold.

In the case of the decaying turbulence, the loop equations represent a closed subset of the Hopf equations, which are still sufficient to generate the dynamics of vorticity.
In this case, the exact solution we have found for the Hopf functional also follows from the integrals of motion, this time, the conservation laws in the loop space.
The loop space Hamiltonian we have derived from the unforced \NS{} equation does not have any potential terms ( those with explicit dependence upon the shape of the loop).
The Schrödinger equation with only kinetic energy in the Hamiltonian conserves the momentum. The corresponding wave function is the plane wave $\exp{\imath \vec p \cdot \vec x}$.
This plane wave is the solution we have found, except the dot product $\vec p \cdot \vec x$ becomes a symplectic form $\oint \vec P(\theta) \cdot d \vec C(\theta)$ in the loop space.
Our momentum $\vec P(\theta, t)$ is not an integral of motion, but simple scaling properties of the pure \NS{} equation lead to the solution with $\vec P(\theta, t) \propto \vec F(\theta)/\sqrt{t}$, with $\vec F(\theta)$ being the integral of motion. 
The rest is a purely technical task: substituting this scaling solution into the \NS{} equation and solving the resulting universal equation for a fixed point $\vec F(\theta)$.

Our solution expresses the probability distribution and expected value for the Wilson loop at any given moment $t$ in terms of an ensemble of fractal loops in complex momentum space. The loop is represented by a polygon with $N \to \infty$ sides.
This statistical system is similar to a one-dimensional Ising ring in an imaginary magnetic field $\imath \beta = \frac{2\imath \pi p}{q}$ and zero coupling constant. 
Some global observables, such as the moments of enstrophy, are calculable for arbitrary $N$ as an analytic function of $N, p, q$, relating it to the Euler totients and similar functions of the number theory.
The continuum limit $N\to \infty$ differs for odd and even $N$, which means this limit does not exist. 
This ambiguity disappears in the grand canonical ensemble.
In this ensemble, the number $N$ of degrees of freedom is not fixed but can also fluctuate, with the weight $\exp{- \mu N}$. These fluctuations smooth out the difference between odd and even ensembles so that the grand canonical ensemble is unambiguous in the continuum limit.
The continuum limit in the grand canonical ensemble corresponds to $\mu \to 0$. In this limit, we compute the partition function \eqref{Zmu} and the expectation value of the energy dissipation \eqref{finalAnswer}. As discussed in the previous section, the singularities at $\mu \to 0$ indicate inconsistencies in the \NS{} equation as an idealization of molecular dynamics.
\subsection{Classical flow and quantum geometry}

Our computations heavily rely on the number theory, particularly Lehmer's multitotients $\varphi_l(q) $, \eqref{new sum}, generalizing \cite{multitotients} the Euler totient function.
What could the number theory have in common with the turbulent flow?
The quantization of parameters of the fixed manifold of decaying turbulence stems from the deep quantum correspondence we have discovered. The statistical distribution of a nonlinear classical \NS{} PDE is exactly related to the wave functional of a linear Schrödinger equation in the loop space. 
The quantization mechanism is the same as in ordinary quantum mechanics: this is a requirement of the periodicity of the solution.
The equivalence of a strong coupling phase of the fluctuating vector field to quantum geometry is a well-known duality phenomenon in gauge theory (the ADS/CFT duality), ringing a bell to the modern theoretical physicist.
In our case, this is a simpler quantum geometry: a fractal curve in complex space.

An expert in the traditional approach to turbulence may wonder why the loop equation's solutions have any relation to the velocity field's statistics in a decaying turbulent flow.
Such questions were raised and answered in the last few decades in the gauge theories, including  QCD\cite{MLDMig86, Mig98Hidden, LoopEqBootstrap, KazakovLooqBootstrap} where the loop equations were derived first \cite{MMEq79, Mig83}. The short answer is that duality only applies to the correlation functions of two theories with different dynamical variables; there is no correspondence between these variables, but the correlation functions are identical.
Mathematical physics sometimes has alternative languages for the same phenomena; examples are the duality between Schrödinger's wave equation and Heisenberg's matrix mechanics, between dynamical triangulation and Liouville theory in 2D quantum gravity. 

Extra complications in the gauge theory are the short-distance singularities related to the infinite number of fluctuating degrees of freedom in quantum field theory. The Wilson loop functionals in coordinate space are singular in the gauge field theory and cannot be multiplicatively renormalized.
Perturbatively, there is no short-distance divergence in the Navier-Stokes equations nor the \NS{} loop dynamics. The Euler equations represent the singular limit, which, as we argued, should be resolved using singular topological solitons regularized by the Burgers vortex.
In the present theory, we keep viscosity constant and do not encounter any singularities in coordinate space. The anomalous dissipation is achieved in our solution via a completely different mechanism: large fluctuations of the fractal curve at $p \ll q$.
However, we have found the singularity from large vorticity fluctuations at any finite viscosity value. It cannot be attributed to the Euler singularities such as line vortexes. Those vortexes are regularized by finite viscosity, unlike our singularities.

This singularity resembles the conjecture by Tao \cite{Tao2016}; however, it has a different meaning. Our singularity displays itself in correlation functions of vorticity, not the local vorticity as a function of coordinates.
Our vorticity is not a smooth field, developing finite-time singularity at some region of physical space, like a vortex line or a vortex sheet; it is a stochastic field with the singularities (discontinuities) distributed all over the physical space with some multi-dimensional distribution. We compute the correlation functions for this distribution from the dual theory of momentum loop dynamics. 
We have singularities in these correlation functions, like QED, though these singularities emerge in the exact solution beyond perturbation theory.

\subsection{Stokes-type functionals and vorticity correlations}

The loop equation describes the gauge invariant sector of the gauge field theory. Therefore, the gauge degrees of freedom are lost in the loop functional. However, the gauge-invariant correlations of the field strength are recoverable from the solutions of the loop equation\cite{MMEq79, Mig83}.
There is no gauge invariance regarding the velocity field in fluid dynamics (though there is such invariance in the Clebsch variables \cite{M23PR}). The longitudinal, i.e., a potential part of the velocity, has a physical meaning  -- it is responsible for pressure and energy pumping. This part is lost in the loop functional but is recoverable from the rotational part (the vorticity) using the Biot-Savart integral.
In the Fourier space, the correlation functions of the velocity field are algebraically related to those of vorticity $ \vec v_k = \frac{\imath \vec k \times \vec \omega_k} {\vec k^2} $. Thus, the general solution for the Wilson loop functional $\Psi[\gamma, C]$ allows computing both vorticity and velocity correlation functions.
We demonstrated that in the last two sections by computing the moments of the enstrophy and resulting anomalous dissipation.
This computation is nonperturbative: it corresponds to the extreme turbulent limit and cannot be expanded in inverse powers of viscosity.

\subsection{Relation of our solution to the weak turbulence}

The solution of the loop equation with finite area derivative, satisfying Bianchi constraint, belongs to the so-called Stokes-type functionals \cite{MMEq79}, the same as the Wilson loop for Gauge theory and fluid dynamics. 
The \NS{} Wilson loop is a case of the Abelian loop functional, with commuting components of the vector field $\vec v$.
As we discussed in detail in \cite{MMEq79, Mig83, M23PR}, any Stokes-type functional $\Psi[\gamma, C]$ satisfying boundary condition at shrunk loop $\Psi[0]=1$, and solving the loop equation can be iterated in the nonlinear term in the \NS{} equations (which iterations would apply at large viscosity). 

The resulting expansion in inverse powers of viscosity (weak turbulence) exactly coincides with the ordinary perturbation expansion of the \NS{} equations for the velocity field, averaged over the distribution of initial data or boundary conditions at infinity.
We have demonstrated in \cite{M93, M23PR} (and also here, in Section \ref{RandomRot}) how the velocity distribution for the random uniform vorticity in the fluid was reproduced by a singular momentum loop $\vec P(\theta)$. 
The solution for $\vec P(\theta)$ in this special fixed point of the loop equation was random complex and had slowly decreasing Fourier coefficients, leading to a discontinuity $\sign(\theta-\theta')$ in a pair correlation function \eqref{Pcorr}. The corresponding Wilson loop was equal to the Stokes-type functional \eqref{InitPsi}.
Using this example, we demonstrated how a discontinuous momentum loop describes the vorticity distribution in the stochastic \NS{} flow. In this example, the vorticity is a global random variable corresponding to a random uniform fluid rotation: a well-known exact solution of the \NS{}  equation. 
This example corresponds to a special fixed point for the loop equation, not general enough to describe the turbulent flow but mathematically ideal as a toy model for the loop technology. It demonstrates how the momentum loop solution sums up all the terms of the $1/\nu$ expansion in the \NS{} equation.  

In our general solution, with the Euler ensemble, the summation of a divergent perturbation expansion occurs at an extreme level, leading to a universal fixed point independent of viscosity.
At a given initial condition, after a finite time, the solution will still depend on viscosity and initial condition.
At large time, though, it will approach our universal fixed manifold and (supposedly, for random initial data) cover it uniformly, according to the Euler ensemble measure. 
The vorticity will become a random variable with a singular distribution in the local limit, suggesting intrinsic inconsistency of the \NS{} equation.

\subsection{Continuum spectrum of anomalous dimensions and multifractality}

We studied the linear perturbations of our fixed manifold, which decay with time as $ t^{-\lambda}$ with the spectrum of $\lambda$ determined by the large $N$ polynomial equation \eqref{spectralEqPolynomial}.
The calculable time decay spectrum at finite $N$ is a nice surprise, but the limit $N \to \8$ remains an unsolved problem. Future investigation will relate it to some "multifractal" properties in the probability distribution emerging after the saddle point computation.
The effective decay spectrum is continuous, meaning the modification of naive scaling laws for decaying turbulence by some powers of the logarithm of time. The analytic result for the large-time asymptotic would require more work, however. Meanwhile, we are involved in the numerical study of this asymptotic behavior \cite{MigBul23}.

\subsection{Conclusion}
\begin{itemize}
    \item The solution for $\vec P(\theta)$ in the loop equation for decaying turbulence, which we have found in this paper, exhibits nonperturbative features, particularly the quantization of parameters. These quantum effects follow from the exact equivalence of the \NS{} statistics to the quantum mechanics in loop space \cite{M23PR}. The most striking nonperturbative feature is the singularity of vorticity distribution in the local limit for a finite viscosity.
    \item  Compared to the other critical phenomena, this theory is amazingly simple: It is not a field theory but a quantum statistical system similar to the one-dimensional Ising ring in the presence of the imaginary quantized magnetic field. The moments of the distribution of vorticity are analytically calculable.
    \item Still, the solution is far from trivial and reveals unexpected relations between turbulence and number theory. 
    The analytic computation of the decay spectrum from the general equation \eqref{resolvent} for the resolvent remains a challenge to the number theory.
    \item This work does not claim a full solution to the decaying turbulence problem. At best, this is the new path to the solution; following this path will require joint efforts of mathematical physicists, number theorists, and computer scientists.
\end{itemize}

\vspace{6pt}
\funding{This 
 research was supported by a Simons Foundation award ID $686282$ at NYU Abu Dhabi.}

\dataavailability{The \Mathematica~notebooks used to verify the equations and compute some functions are available for download in~\cite{DecayTurb23}.} 

\acknowledgments{I benefited from discussions of this theory with Sasha Polyakov, K.R. Sreenivasan, Peter Sarnak, Tristan Buckmaster, Greg Eyink, Luca Moriconi, Vladimir Kazakov, Kartik Iyer, and Maxim Bulatov.
Maxim helped me compute the number theory function $S(q)$, used in the enstrophy's local limit (to be published in~\cite{MigBul23}).}

\conflictsofinterest{The authors declare no conflict of interest.'' 
} 
\appendixtitles{yes}
\appendixstart
\appendix

\section{Euler Averages as Multitotient Functions}\label{Totients}

The Euler ensemble expectation value of $\cot^2\left(\frac{\pi p}{q}\right)$ can be reduced to the prime numbers as follows.
We start with an unconstrained sum, which is elementary \cite{DecayTurb23}. 
\begin{equation}
    \sum_{n=1}^{m-1} \cot^2\left(\frac{\pi n}{m}\right) =  \frac{1}{3} m^2 - m + \frac{2}{3} 
\end{equation}

Let us consider a sum $G[F,m]$ of arbitrary function $F(n/m)$ constrained to the coprime $n,m$.
In our case 
\begin{eqnarray}
    F(x) = 
    \begin{cases}
        & 0 \text{ if x =1}\\
        & \cot^2\left(\pi x\right) \text{ otherwise}\\
    \end{cases}
\end{eqnarray}

Such sum satisfies  the following equation (with $p_{1} < p_{2}\dots <p_{S}$ denoting $S$ ordered prime factors of $m$ and $(n,m)$ denoting coprime $n,m$).
\begin{eqnarray}
&& m = \prod_{s=1}^S p^{\alpha_s}_{s};\\
&&G[F,m] =\sum_{\substack{n=1\\(n,m)=1}}^{m} F(n/m);\\
&&H(m) = \sum_{n=1}^{m} F(n/m);\\
\label{basicEq}
&&G[F,m] = H(m) +\sum_{s=1}^{S} (-1)^s \sum_{0<l_1<l_2\dots<l_s\leq S}H\left(\frac{m}{p_{l_1} p_{l_2}\dots p_{l_s}}\right);
\end{eqnarray}

Let us go into detail. Consider the first term for particular $l_1 = l, p(l_1) = p$
\begin{eqnarray}
   && H(m) -  H\left(\frac{m}{p}\right) = \sum_{n=1}^{m} F(n/m) -  \left.\sum_{n'=1}^{m'} F(n'/m')\right|_{m' = m/p}=\nonumber\\
    &&\sum_{n=1}^{m} F(n/m) -  \sum_{\substack{n=1\\n\Mod{p}= 0}}^{m} F(n/m)
\end{eqnarray}

We observe that the second term removes from the total sum $\sum_{n=1}^m $ in the first term all the terms with $n\Mod{p}= 0$.
In the same way, the other terms in the sum $\sum_l H\left(\frac{m}{p_l}\right)$ remove all the terms in the first sum with $(n,m) = p_l$.
However, there are terms in $\sum_{n=1}^{m} F(n/m)$ like $n = p_1 p_2$, which are proportional to two prime factors $p_1, p_2$, and we removed these terms twice, once in the term $-H\left(\frac{m}{p_1}\right)$ and the second time in $-H\left(\frac{m}{p_2}\right)$.
So, we have to add them back, with $+1$ sign for each pair $p_i, p_j$. This addition provides the next term with double sum $ \sum_{0 < l_1 < l_2 \le S}$.

In general, this formula is a particular case of the inclusion-exclusion principle \cite{IncExc}.
As the basic equation \eqref{basicEq} is a linear functional of $H$, we can solve this equation separately for $H_l(m) = m^l$, and then by adding these solutions with proper coefficients, we get the solution for our particular $H(m) = \tt -m + \ot m^2$.
Let us start with the simplest case, $H_1(m) =m$. 
The solution is the Euler $\varphi(m)$. Here is how the equation is satisfied:
\begin{eqnarray}
     && G_1(m) = m + m \sum_{s=1}^S (-1)^s \sum_{0<l_1<l_2\dots<l_s\leq S}\frac{1}{p_{l_1} p_{l_2}\dots p_{l_s}}= \nonumber\\
     && m \prod_{l=1}^S\left(1 - \frac{1}{p_l}\right)
\end{eqnarray}

The next case, $G_2(m)$ is processed the same way, with the result
\begin{eqnarray}
     && G_2(m) =  m^2\sum_{s=0}^L(-1)^s \sum_{l_1<l_2\dots<l_s}\frac{1}{p^2(l_1)p^2(l_2)\dots p^2(l_s)} =\nonumber\\
     &&m^2 \prod_{l=1}^L\left(1 - \frac{1}{p_l^2}\right)
\end{eqnarray}

Finally, the function $G_0(m)$ 
\begin{eqnarray}
    G_0(m)= \sum_{s=0}^S (-1)^s \sum_{l_1<l_2\dots<l_s}1 = \sum_{s=0}^S (-1)^s \Binom{S}{s} = (1-1)^S = 0
\end{eqnarray}

Putting all together
\begin{eqnarray}
    && S(m) =\sum_{\substack{n=1\\(n,m)}}^{m-1} \cot^2\left(\frac{\pi n}{m}\right) =\frac{1}{3} \varphi_2(m) - \varphi_1(m) ;\\
    && \varphi_l(m) = m^l \prod_{p|m}\left(1 - \frac{1}{p^l}\right) ;
\end{eqnarray}

These multitotients $\varphi_l(m)$ were introduced by Lehmer in $1900$ \cite{multitotients}.
The asymptotic behavior of the multitotient summators was computed in that paper:
\begin{eqnarray}
    \sum_{m=2}^N\varphi_l(m)  \to \frac{ N^{l+1}}{ (l +1) \zeta(l+1)}
\end{eqnarray}

We also need similar sum rules for higher powers of $\cot(\beta)$. 
\begin{eqnarray}
    S(n,m) = \sum_{\substack{p=1\\(p,m)=1}}^{m-1}\cot^{2n}\left(\frac{\pi p}{m}\right);
\end{eqnarray}

This sum belongs to the general category of the sums $G[F_n,m]$ with 
\begin{eqnarray}
    F_n(x) = 
    \begin{cases}
        & 0 \text{ if x =1}\\
        & \cot^{2 n}\left(\pi x\right) \text{ otherwise}\\
    \end{cases}
\end{eqnarray}

Repeating the above arguments, we only need to know the polynomial expansion of the unconstrained sum of the $nth$ power of cotangent.
This sum was computed in \cite{Franke2021}. The expression in that paper contained a $n$-fold sum and thus was hard to use in any analytic or numerical computation.
We reduced this multiple sum to the following linear  recurrent equation (with $B_k$ being the Bernoulli coefficients)
\begin{eqnarray}
&&\text{BernSum}(0,0) =1;\\
&&\text{BernSum}(n,m) = 0 \text{ if } n < m;\\
&& \text{BernSum}(n,m)=\sum _{j_1=0}^m \sum _{j_2=0}^{m-j_1} \frac{B_{2 j_1} B_{2 j_2} \text{BernSum}(n-1,m-j_1-j_2)}{(2 j_1)! (2 j_2)!};
\end{eqnarray}

The coefficients of the cotangent sum are related to these coefficients $\text{BernSum}$ as follows
\begin{eqnarray}
   &&H[F_n,m] = (-1)^n m - (-4)^n  \sum _{j=0}^n \frac{B_{2 j} m^{2 j} \text{BernSum}(n,n-j)}{(2 j)!};
\end{eqnarray}

This recurrent equation is easily solved for any finite $n$ and, being linear, can also be analyzed in the limit of large $n$.
Here are the first four sums $H[F_n,m], n = 0, 1,2,3$
\begin{eqnarray}
    \left(
\begin{array}{c}
m \\
 \frac{1}{3} (m-2) (m-1) \\
 \frac{1}{45} \left(m \left(m^3-20 m+45\right)-23\right) \\
 \frac{1}{945} \left(m \left(2 m^5-42 m^3+462 m-945\right)+396\right) \\
\end{array}
\right)
\end{eqnarray}

The last step is to replace the powers of $m$ with corresponding multitotients, according to the above theory
\begin{eqnarray}
     &&S(n,m) = (-1)^n \varphi(m) - (-4)^n  \sum _{j=1}^n \frac{B_{2 j} \varphi_{2 j}(m)\text{BernSum}(n,n-j)}{(2 j)!};
\end{eqnarray}

Note that the term without a power of $m$ dropped as $\phi_0(m) =0$.
Here are the first four sums
\begin{eqnarray}
    \left(
\begin{array}{c}
\varphi_1(m)\\
 -\varphi_1(m) +\frac{1}{3} \varphi_2(m)\\
 \varphi_1(m)+\frac{1}{45} (\varphi_4(m)-20 \varphi_2(m)) \\
 -\varphi_1(m) +\frac{1}{945} (462 \varphi_2(m)-42 \varphi_4(m)+2 \varphi_6(m)) \\
\end{array}
\right)
\end{eqnarray}

The \Mathematica does not know how to simplify the sums like $H[F_n,m], S(n,m)$  for $ n >1$, but we can use it to numerically compute these sums and compare them with our totient solution.
Here is an example for $S(5, 1000)$
\begin{eqnarray}
    \begin{array}{l|ll}
 \text{} & \text{rational} & \text{numerical} \\
\hline
\text{\Mathematica{}} & None                                 & 2.1356217511751929448\times 10^{25} \\
\text{totients} & \frac{4036325109694867136069546800}{189}   & 2.1356217511613053630\times 10^{25} \\
\end{array}
\end{eqnarray}

Only the first $10$ digits out of the requested $20$ came out correct with numerical evaluation.
The real advantage of the exact solution is that its computational complexity grows only logarithmically with $m$, at least for the achievable values $ m \sim 10^7$ where the prime factorization does not present a problem.
The original definition with the sum over comprime $(j,m)$ has a linear complexity.


\begin{adjustwidth}{-\extralength}{0cm}

\reftitle{References}

\PublishersNote{}
\end{adjustwidth}

\end{document}